\RequirePackage{fix-cm}
\documentclass[twocolumn,epjc3]{svjour3}  
\smartqed  
\RequirePackage{graphicx}
\RequirePackage{latexsym}
\RequirePackage{amsmath,amsfonts}
\RequirePackage[colorlinks,citecolor=blue,urlcolor=blue,linkcolor=blue]{hyperref}
\journalname{Eur. Phys. J. C}
\begin{document}

\title{Jeans analysis in energy-momentum-squared gravity}


\author{Ali Kazemi\thanksref{addr1}
        \and
        Mahmood Roshan\thanksref{e2,addr2,addr1}
        \and
        Ivan De Martino\thanksref{addr3,addr3-1,addr4}
        \and
        Mariafelicia  De Laurentis\thanksref{addr4,addr5,addr6}        
}

\thankstext{e2}{e-mail: mroshan@um.ac.ir}


\institute{
School of Astronomy, Institute for Research in Fundamental Sciences (IPM), 19395-5531, Tehran, Iran \label{addr1}
\and
Department of Physics, Faculty of Science, Ferdowsi University of Mashhad P.O. Box 1436, Mashhad, Iran\label{addr2}
\and
Donostia International Physics Center (DIPC), 20018 Donostia-San Sebastian (Gipuzkoa) Spain\label{addr3}
\and
Dipartimento di Fisica, Universit\`a di Torino,  Via P. Giuria 1, I-10125 Torino, Italy\label{addr3-1}
\and
Istituto Nazionale di Fisica Nucleare (INFN), Sezione di Torino, Via P. Giuria 1, I-10125 Torino, Italy\label{addr4}
\and
Dipartimento di Fisica "E. Pancini", Universit\'a di Napoli "Federico II", Compl. Univ. di Monte S. Angelo, Edificio G, Via Cinthia, I-80126, Napoli, Italy\label{addr5}
\and
Lab.Theor.Cosmology,Tomsk State University of Control Systems and Radioelectronics(TUSUR), 634050 Tomsk, Russia\label{addr6}
}

\date{Received: date / Accepted: date}

\maketitle

\begin{abstract}
In this paper, we study the Jeans analysis in the context of energy-momentum-squared gravity (EMSG). More specifically we find the new Jeans mass for non-rotating infinite mediums as the smallest mass scale for local perturbations that can be stable against its own gravity. Furthermore, for rotating mediums, specifically for rotating thin disks in the context of EMSG, we find a new Toomre-like criterion for the local gravitational stability. 
Finally, the results are applied to a hyper-massive neutron star, as an astrophysical system. Using a simplified toy model we have shown that, for a positive (negative) value of the EMSG parameter $\alpha$, the system is stable (unstable) in a wide range of $\alpha$. On the other hand, no observational evidence has been reported on the existence of local fragmentation in HMNS. Naturally, this means that EMSG with positive $\alpha$ is more acceptable from the physical point of view.
\end{abstract}

\section{Introduction}
Recent statistical analyses of astrophysical and cosmological datasets have once again confirmed the concordance $\Lambda$CDM model \cite{Blake2011,Suzuki2012,Hinshaw2013,Planck18}. Despite its successes, the model shows some shortcomings. On one side, the fundamental nature of the two most important energy density components, namely Dark Energy (DE) and Dark Matter (DM), is still unknown \cite{bertone2005,caldwell2009,Feng2010,atrio2016}. Many candidates have been proposed without being able to solve the puzzle \cite{PR03,Pad03,D+05,CTTC06,Caldwell02,PR88,RP88,SS00,Schive2014,Capolupo2016,Capolupo2017,Kleidis2011,Kleidis2015,Kleidis2017,demartino2017b,demartino2018,capolupo2019}. On the other hand, it is well known that General Relativity (GR) is not a Quantum Theory of gravity and it cannot provide a description of the Universe at the quantum scales needed to solve the fine-tuning of initial conditions \cite{Misner1970,Itzykson1980,Isham1981}. As a consequence, many modified theories of the gravity have been proposed to solve the puzzle \cite{Faraoni2009,darkmetric,Nojiri2011,PhysRept,Annalen2012,idm2015,Nojiri2017,Nojiri:2006ri,Cai2016}. Nevertheless, having alternative explanations demands to test the modified gravity models and other basic tenets of the $\Lambda$CDM cosmology at all scales, both in the strong and weak field regime. In particular, let us remember that the constraints at the Solar System scale must be matched by any theory of gravity under consideration \cite{Will93,Stairs2003,Everitt2011}. 

Here, we will compute the weak field limit of Energy-Momentum-Squared-Gravity (EMSG), recently introduced by \cite{us,katirci2014f}, to study the collapse of a self-gravitating system. The main idea behind EMSG is to resolve the Big Bang singularity in a non-quantum description. It is important to mention that, GR inherently leads to the singularity in the early universe. On the other hand, as already mentioned, in the early universe, i.e., at the Planck scale, the quantum gravity effects play an important role. Therefore GR predicts space-time singularity in a physical situation in which its viability is seriously doubted. EMSG's action functional is obtained by adding scalar terms proportional to $T_{\mu\nu}T^{\mu\nu}$ (where $T_{\mu\nu}$ is the energy-momentum tensor) to the Einstein-Hilbert action, and it leads to interesting cosmological behaviours This kind of corrections, naturally induce squared contributing terms like $\rho^2$, $p^2$ and $\rho p$ to the Friedman equations governing the background cosmological evolution. Where $\rho$ and $p$ are the energy density and pressure of the cosmic fluid. As a consequence, there are bouncing cosmological solutions in this model, and the cosmic scale factor cannot be smaller than a minimal length scale. In other words, there is a finite maximum energy density. This directly means that EMSG can prevent the Big Bang singularity in a completely non-quantum way. More importantly, EMSG does not alter the cosmological evolution. Its only effect is to resolve the singularity (for more details we refer the reader to \cite{us}). However, it is should be stressed that EMSG's effects can appear also in the stellar configurations. For example, it is shown in \cite{nari} that EMSG can lead to more massive neutron stars than in GR. This fact is satisfactory in the sense that there are difficulties in GR for explaining the internal structure of massive neutron stars, especially their high mass, using ordinary equations of state (for more details see \cite{Demorest:2010bx} and \cite{Antoniadis:2013pzd}).

The study of the collapse of a self-gravitating system is somehow the first test to do to probe any modified theory of gravity. Indeed, Jeans Instability for a spherically symmetric self-gravitating systems causes the collapse of a gas cloud under the gravitational force giving rise to the formation of self-gravitating structure such as stars and galaxies among the others \cite{Binney}. For stability, the cloud must be in hydrostatic equilibrium, and this physical condition holds only on certain scales determined by the so-called Jeans length, $\lambda_J^2 = \frac{c_s^2}{2 G\rho}$ where $c_s$ is the sound speed, $G$ is the Newton's gravitational coupling constant and $\rho$ is the matter density.
All perturbations having wavelengths larger than it are unstable. On the contrary, smaller wavelengths are stable. Since such a scale is strongly dependent by the underlying theory of gravity, it has been used to probe several modified theories of gravity \cite{Capozziello2012,idm2017a,Arbuzova2014,Roshan2014}. 

Besides the stability criteria for spherically symmetric perturbations, \cite{Toomre1964} investigated the stability condition of all local axisymmetric perturbations introducing the dimensionless parameter $Q=\frac{c_s \kappa}{\pi G \Sigma}$, where $\kappa$ is the epicyclic frequency and $\Sigma$ is the surface density of the system. Thus, any cloud or disk is stable if the condition $Q>1$ holds. As it was for Jeans instability, it has been shown that also the Toomre's criterion can be used to check the validity of several modified theories of gravity \cite{Roshan2015a,Roshan2015b,RoshanEiBI}. 
Generalizing both criteria for the local stability in the framework of EMSG could provide a very remarkable tool to describe the dynamics of
self-gravitating system such as the collapse of spherical clouds, the collapse of massive star into Black Hole and/or the accretion disks around a massive object, leading to new results that could potentially be used to retain/rule out the theory.

The paper is organized as follows: In Sec. \ref{fieldE} we briefly introduce the EMSG and derive its field equations. In Sec. \ref{wf} we perform the weak field limit of EMSG. In particular, we write down the modified Poisson's equation. In Sec. \ref{Sect:hydroeq}, we give the modified continuity and Euler equation for EMSG. In Sec. \ref{jeans_EMSG} and \ref{tomre_emsg}, we compute the Jeans's length and the Toomre parameter for EMSG, respectively. In both cases we compute and analyze the dispersion relation particularizing our calculation to specific cases of the EMSG. In Sec. \ref{disk_emsg}, we analyze the stability of an exponential disk to recover the Toomre's criteria and, then, in Sec. \ref{applications} we apply our calculations to the case of Hyper Massive Neutron Stars (HMNS). Finally, in Sec. \ref{conclusions} we summarize our results and conclusions.

\section{Field equations of EMSG}
\label{fieldE}
As in any other theory of gravity, the starting point of the EMSG is the action
\begin{equation}
S=\frac{1}{2\gamma}\int f(R,\mathbf{T}^2)\sqrt{-g}\,d^4 x+\int \mathcal{L}_M \sqrt{-g}\,d^4x\,,
\end{equation}
where $\gamma=8\pi G/c^4$, $G$ Newton's constant, $c$ is the speed of light, $\sqrt{-g}$ is the determinant of metric tensor, $\mathcal{L}_M$ is the matter Lagrangian density, $\mathbf{T}^2=T_{\mu\nu}T^{\mu\nu}$, and $T_{\mu\nu}$ is the energy-momentum tensor \cite{us}. Notice that, we use the metric signature $(-,+,+,+)$.  Working in the metric formulation of the theory, it is straightforward to vary the action with respect to the metric and find the following field equations
\begin{equation}
f_R R_{\mu\nu}-\frac{1}{2} g_{\mu\nu }f=\gamma T_{\mu\nu}-\Big[f_Q \theta_{\mu\nu}+(g_{\mu\nu}\qed-\nabla_{\mu}\nabla_{\nu})f_R\Big]\,,
\label{fe1}
\end{equation}
where $f=f(R,\mathbf{T}^2)$, $f_R=\partial f/\partial R$ and $f_Q=\partial f/\partial Q$, the $\qed$ is the usual d'Alembert operator, and for simplicity in notation we have defined $Q\equiv\mathbf{T}^2$.  On the other hand, the tensor $\theta_{\mu\nu}$ is defined as the variation of $Q$ with respect to the metric tensor, namely $\theta_{\mu\nu}=\delta Q/\delta g_{\mu\nu}$. For a perfect fluid the energy-momentum tensor and $\theta_{\mu\nu}$ are written as follows (for more details see \cite{barrow,akarsu2018cosmic})
\begin{equation}
T_{\mu\nu}=(\rho+\frac{p}{c^2})u_{\mu}u_{\nu}+g_{\mu\nu}p\,,
\end{equation}
and 
\begin{eqnarray}
&\theta_{\mu\nu}=-(\rho^2 c^2+4p \rho+3 \frac{p^2}{c^2})u_{\mu}u_{\nu}\\
& Q=\rho^2 c^4+3 p^2
\end{eqnarray}
where $\rho$ and $p$ are the energy density and pressure of the perfect fluid, respectively. Moreover $u^{\mu}$ is the four velocity of the fluid. Before moving on to discuss the weak field limit of the theory, let us take the trace of field Eqs.~ (\ref{fe1}). The result is written as
\begin{equation}
f_R R-2 f=\gamma T-(f_Q \theta+3\qed f_R)\,,
\label{fe2}
\end{equation}
where $\theta=g^{\mu\nu}\theta_{\mu\nu}$. Now, we have all the equations needed to perform the weak field limit of EMSG.

\section{Weak field limit of EMSG}\label{wf}

Let us  compute the first order perturbations of the field equations around the Minkowski space time in order to find the governing equations for the Newtonian self-gravitating disk in the context of EMSG. To do so we write the line element in the Cartesian coordinate $(ct,x,y,z)$ using the perturbed metric, i.e. $g_{\mu\nu}=\eta_{\mu\nu}+h_{\mu\nu}$ where $|h_{\mu\nu}|\ll |g_{\mu\nu}|$, as follows
\begin{equation}
ds^2=-(1+\frac{2\,\Phi}{c^2})c^2dt^2+(1+\frac{2\,\Psi}{c^2})(dx^2+dy^2+dz^2)\,.
\label{line_element}
\end{equation}

The corresponding first order perturbations in other quantities can be written as
\begin{eqnarray}
&& \label{per1a} Q=Q^0+\delta Q\,,\\
&& \label{per1b} R=R^0+\delta R\,,\\
&& \label{per1c} \theta_{\mu\nu}=\theta^0_{\mu\nu}+\delta\theta_{\mu\nu}\,,\\
&& \label{per1d} f=f^0+f^0_R \delta R+f^0_Q \delta Q\,,\\
&& \label{per1e} f_R=f_R^0+f_{RR}^0\delta R +f_{RQ}^0 \delta Q\,,\\
&& \label{per1f} f_Q=f_Q^0+f_{RQ}^0\delta R +f_{QQ}^0 \delta Q\,,
\end{eqnarray}
where the suffix "$0$" indicates the background quantities, and $f_{XY}=\partial^2 f/\partial X\partial Y$. For the Minkowski background we have $T_{\mu\nu}^0=0$ and consequently $Q^0=0$. We assume that the function $f(R,Q)$ is chosen in a way that if $T_{\mu\nu}^0=0$ in the background then the background Ricci scalar $R^0=0$ and $f^0=f(0,0)=0$. In this case, using the definitions of $T_{\mu\nu}$ and $\theta_{\mu\nu}$, it is straightforward to verify that
\begin{eqnarray}
&&  \label{iv1} \delta T_{\mu\nu}\simeq \rho u^0_{\mu}u^0_{\nu}\,,\\
&&\label{iv2}\delta\theta_{\mu\nu}\simeq -\rho^2 c^2 u^0_{\mu}u^0_{\nu}\,,
\end{eqnarray}
where commonly we have assumed that in the weak field limit $p/\rho c^2 \ll 1$. Furthermore one should note that the background velocity four-vector is given by $u^0_{\mu}=(-c,0,0,0)$
Now, let us substitute perturbed quantities in  given Eqs.~(\ref{per1a})-(\ref{per1f}) into Eqs.~ (\ref{fe1}) and (\ref{fe2}). Keeping only the first order terms, Eq. (\ref{fe2}) takes the following form
\begin{equation}
f^0_{RR}\qed \delta R+f^0_{RQ}\qed \delta Q=\frac{\gamma}{3}\delta T-\frac{f^0_Q}{3}\delta \theta+\frac{f^0_R}{3}\delta R+\frac{2f^0_{Q}}{3}\delta Q\,,
\label{per2}
\end{equation}
where $\delta T$ and $\delta \theta$ are perturbations in $T$ and $\theta$ respectively. Hereafter, for brevity in notation and prevent confusion with temporal components of the tensors, we drop the "$0$" suffix. Now we use Eq. (\ref{per2}) to linearize the time-time component of the field Eq.~ (\ref{fe1}) as follows
\begin{equation}
\delta R^0_0=\frac{\gamma}{f_R}\Big(\delta T^0_0-\frac{1}{3}\delta T\Big)-\frac{f_Q}{f_R}\Big(\delta\theta_0^0-\frac{\delta\theta}{3}+\frac{\delta Q}{6}\Big)+\frac{\delta R}{6}\,,
\label{per4}
\end{equation}
using a standard gauge it is straightforward to show that $\delta R^0_0=-\nabla^2 \Phi/c^2$. 
To be precise, the standard gauge is also commonly called standard gauge of post-Newtonian theory (for more details see Sec. 8.3.7 in \cite{Poisson2013}). In particular, this gauge condition allow us to simplify the perturbed field equations at first order. Using this gauge condition we obtain two results: firstly we eliminate the higher order time derivatives of metric tensor, and secondly the Poisson equations can be solved more easily.

On the other hand, in principle, we can consider the perturbed Ricci scalar as a function of $\rho$ and $p$, i.e. $\delta R=\delta R(\rho,p)$. To see this fact more clearly, let us conveniently assume that $f_{RQ}=0$. We will use this assumption everywhere in this paper. This means that we only deal with models that can be recast in the following form 
\begin{equation}\label{eq:fRQform}
f(R,Q)=f_1(R)+f_2(Q)\,.       
\end{equation}
In this case, keeping in mind that $\qed \delta R=\nabla^2 \delta R$, we rewrite Eq. (\ref{per2}) as
\begin{equation}
\nabla^2 \delta R-\mathcal{M}^2 \delta R= H(\rho, p)\,,
\label{per3}
\end{equation}
where the mass $\mathcal{M}^2$ is defined as
\begin{equation}
\mathcal{M}^2=\frac{f_R}{3 f_{RR}}\,,
\label{M2}
\end{equation}
and the function $H$ is
\begin{equation}
H(\rho,p)=\frac{1}{3f_{RR}}\Big[\gamma \delta T-f_Q(\delta \theta-2 \delta Q)\Big]\,.
\end{equation}
It should be noted that it is natural to expect that $f_R$ in the background is unity. However for the sake of completeness we keep it as a free parameter in the calculations. Furthermore let us define new parameter $\alpha$ as 
\begin{equation}
\alpha=\frac{f_Q}{\gamma}\,.
\end{equation}
Consequently, for the general form for $f(R,Q)$ in Eq.~(\ref{eq:fRQform}) with $f_{RR}\neq 0$, we can integrate Eq.~(\ref{per3}) to obtain $\delta R$ in terms of $\rho$ and $p$
\begin{equation}
\delta R=\chi\int\frac{e^{- \mathcal{M} |\mathbf{r}-\mathbf{r}'|}}{|\mathbf{r}-\mathbf{r}'|}\Big[- \delta T(\mathbf{r}')+\alpha(\delta\theta(\mathbf{r}')-2\delta Q(\mathbf{r}'))\Big]d^3 \mathbf{r}'\,,
\label{per0}
\end{equation}
where for convenience, we have defined $\chi\equiv\frac{\gamma \mathcal{M}^2}{4\pi f_R}.$ 

Therefore, using Eq.~(\ref{per4}), the modified version of Poisson's equation in EMSG can be written as
\begin{equation}
\nabla^2 \Phi=\frac{\gamma\, c^4}{2}\tilde{\rho}\,,
\label{per5}
\end{equation}
where we have defined the density $\tilde{\rho}$ as
\begin{equation}
\tilde{\rho}=-\frac{2}{f_R}\Big(\delta T^0_0-\frac{1}{3}\delta T\Big)+\frac{2\alpha}{  f_R}\Big(\delta\theta_0^0-\frac{\delta\theta}{3}+\frac{\delta Q}{6}\Big)-\frac{\delta R}{3\gamma\,c^2}\,.\label{per6}
\end{equation}
On the other hand by using Eqs.~(\ref{iv1}) and (\ref{iv2}) we have
\begin{equation}
\delta T^0_0\simeq-\rho c^2, ~~~~\delta T\simeq-\rho c^2,~~~\delta Q=\delta \theta=\delta \theta^0_0\simeq\rho^2 c^4\,,
\label{per8} 
\end{equation} 
where we have applied the condition $p\ll \rho c^2$ in the weak field limit. 
In GR we have $f_Q=0$. Moreover in this case we have $\delta R=-\gamma \delta T$. Consequently it is easy to show that $\tilde{\rho}=\rho$, and Eq.~(\ref{per5}) recovers the standard Poisson's equation. For another special case, the EMSG model studied in \cite{us} is given by $f(R,Q)=R-\eta \mathbf{T}^2=R-\eta Q$. For this model we have $f_{RR}=0$, $f_R=1$, and $f_Q=-\eta=\alpha\gamma$. 
Moreover from Eq.~(\ref{per2}), one may simply verify that $\delta R=-\gamma (\delta T+\alpha(2\delta Q-\delta \theta))$. Therefore Eq.~(\ref{per6}) gives
\begin{equation}
\tilde{\rho}=\rho (1+2\alpha\rho c^2)\,,
\label{per7}
\end{equation}
in this special case, the effects of EMSG can be included in the effective density and pressure defined as, see \cite{nari} 
\begin{eqnarray}
&& \rho_{\rm{eff}}
=\rho+\frac{\alpha c^2}{2}\left(8\rho \frac{p}{c^2}+\rho^2+3\frac{p^2}{c^4} \right)\,,
\label{effectr}
\\
&& p_{\rm{eff}}
=p+\frac{\alpha c^4}{2}\left(\rho^2+3\frac{p^2}{c^4} \right)\,,
\label{effectp}
\end{eqnarray}
More specifically, it has been shown in \cite{nari} that the governing equations of EMSG, are completely similar to GR and the only difference is that $\rho$ and $p$ are replaced with $\rho_{\rm{eff}}$ and $p_{\rm{eff}}$. In this case, in the weak field limit we can rewrite Eq.~(\ref{per7}) as $\tilde{\rho}=\rho_{\rm{eff}}+3p_{\rm{eff}}/c^2$. In other words the Poisson's equation, as one may expect, takes the following form
\begin{equation}
\nabla^2 \Phi=\frac{\gamma c^4}{2}(\rho_{\rm{eff}}+3\frac{p_{\rm{eff}}}{c^2})\,.
\end{equation}
This is similar to the corresponding equation in GR, where we take into account pressure as a source for gravity (see \cite{ehlers} for more details).

Now before moving on to discuss the Euler equation, let us summarize the weak field limit and write the modified Poisson's equation for two different categories, namely EMSG models
with $f_{RR}=0$ and $f_{RR}\neq 0$. For the first case, using Eqs.~(\ref{per2}) and (\ref{per5})-(\ref{per8}), we arrive at
\begin{equation}
\nabla^2 \Phi=\frac{\gamma \, c^4}{2 f_R}\Big(\rho+2\alpha\rho^2 c^2\Big)\,,
\label{n1}
\end{equation}
and similarly for the second case, using Eqs.~(\ref{per2}) and (\ref{per0})-(\ref{per8}), we find a more complicated Poisson's equation

\begin{eqnarray}
\nonumber\nabla^2 && \Phi=\frac{\gamma c^4}{6 f_R}\Big[4\rho+5\alpha \rho^2 c^2\\
&& -\frac{\mathcal{M}^2}{4\pi}\int\frac{e^{-\mathcal{M}|\mathbf{r}-\mathbf{r}'|}}{|\mathbf{r}-\mathbf{r}'|}\Big(\rho(\mathbf{r}')-\alpha\rho^2(\mathbf{r}') c^2\Big)d^3 r'\Big]\,.
\label{poisson}
\end{eqnarray}

\section{Hydrodynamics equations in weak-field limit}\label{Sect:hydroeq}
To find the Newtonian limit of the hydrodynamics equations, one can take the covariant derivative of the field Eqs.~ (\ref{fe1}) as below
\begin{eqnarray}
\nonumber\left( \nabla^{\mu} f_R\right)R_{\mu\nu} + && f_R\left( \nabla^{\mu}R_{\mu\nu}\right) -\frac{1}{2}g_{\mu\nu}\left( \nabla^{\mu}f\right)=\gamma \nabla^{\mu}\left( T_{\mu\nu}\right)\\
&& +\left(\qed\nabla_{\nu}-\nabla_{\nu}\qed \right) f_R-\nabla^{\mu}\left(f_Q \theta_{\mu\nu} \right).
\label{Div}
\end{eqnarray}
To simplify the third term in the above equation, we recall that
$f(R,Q)=f_1(R)+f_2(Q)$.
Therefore one can easily verify that
\begin{equation}
g_{\mu\nu}\nabla^{\mu}f=g_{\mu\nu}\left(f_R\nabla^{\mu}R+f_Q\nabla^{\mu}Q \right).
\end{equation}
Also, the fifth term can be simplified as below (\cite{Koivisto})
\begin{equation}
\left(\qed\nabla_{\nu}-\nabla_{\nu}\qed \right) f_R=R_{\mu\nu}\nabla^{\mu}f_R.
\end{equation}
Using the Bianchi identity, and after some manipulations, one can find the perturbed form of Eq.~(\ref{Div}) as
\begin{equation}
\nabla^{\mu}\left( \delta T_{\mu\nu}\right) =\alpha\left(\nabla^{\mu}\left(\delta\theta_{\mu\nu} \right) -\frac{1}{2}\eta_{\mu\nu}\nabla^{\mu}\left( \delta Q\right)  \right),
\end{equation}
where $\delta Q=\delta T_{\mu\nu}\delta T^{\mu\nu}$. Note that, the background quantities are shown without the ``0" index here. To achieve the hydrodynamics equations in the Newtonian limit, one can ignore the terms containing the pressure compared with the similar terms containing the density. In fact, the pressure plays role in the relativistic situations, which are not, of course, of interest in this study. 

Let us look at the order of magnitudes. What we have assumed is: firstly, as mentioned before, the pressure can be ignored comparing with the density in our background system. Secondly, the gravitational field assumed to be weak. And finally, the velocities inside the background are slow. Using a small parameter $\epsilon$, these assumptions can read
\begin{equation}
\frac{p}{\rho c^2}\simeq \frac{v^2}{c^2}\simeq \frac{\Phi}{c^2} \propto \epsilon^2
\end{equation}
On the other hand, considering the Newtonian form of the Euler's equation, one can see that $\partial\vec{v}/\partial t\simeq (\vec{v}\cdot\nabla)\vec{v}\simeq\nabla\Phi$ and, therefore
\begin{equation}
\frac{\partial}{\partial t}\simeq \vec{v}\cdot \nabla \propto \epsilon
\end{equation}
Moreover, remembering the smallness of $\alpha$, some terms containing a multiplication of $\alpha$ and $\epsilon$ should be treated carefully. In fact, one can consider the same order of magnitude for these parameters, and keep the terms only up to $O(\epsilon^2)$. It is worth mentioning that, the parameter $\epsilon$ is only a useful gadget to track the order of magnitudes. After finding the hydrodynamics equation, one can truly assume that $\epsilon\rightarrow1$. 

Now, keeping in mind that  $u^{\mu}=(c,\vec{v})$, 
and by finding $\delta T_{\mu\nu}$, $\delta\theta_{\mu\nu}$, and $\delta Q$, one can easily decompose the Eq.~(\ref{Div}) to the temporal and spacial components. It is worth mentioning that, during the simplification of the components, the terms containing $\alpha \epsilon^2$ can be ignored. Furthermore, the terms including a temporal derivative multiplied by $\epsilon^2$ or $\alpha\epsilon$ can be ignored.  
After some manipulations one can show that the $t$-component of the Eq.~(\ref{Div}), can be written as
\begin{eqnarray}
\nonumber(1+\alpha\rho c^2)\frac{\partial\rho}{\partial t} +&&\epsilon\left\lbrace (1+\alpha\rho c^2)\rho\nabla\cdot\vec{v}+\right.\\ &&\left.(1+2\alpha\rho c^2)\vec{v}\cdot\nabla\rho\right\rbrace =0,
\end{eqnarray}
then one may simply rewrite this equation as
\begin{equation}
\frac{\partial\rho}{\partial t} +\nabla\cdot (\rho \vec{v})= - \frac{\alpha \rho c^2}{1+\alpha \rho c^2} \vec{v}\cdot \nabla \rho
\label{cn2}
\end{equation}
this equation, is the continuity equation in the weak-field limit of the $f(R,Q)$ gravity. 
It is not difficult to show that, in terms of the effective quantities defined in Eqs.~(\ref{effectr}) and (\ref{effectp}), the continuity equation in the Newtonian limit can be written as
\begin{equation}
\frac{\partial\rho_{\rm{eff}}}{\partial t}+\nabla\cdot\biggl[\left(\rho_{\rm{eff}}+\frac{p_{\rm{eff}}}{c^2} \right) \vec{v}\biggr]=0.
\end{equation}

Now, from the spacial components of Eq.~(\ref{Div}) and using Eq.~(\ref{cn2}) after some manipulations one can find the Euler's equation in the weak-field limit of this theory. To do so we obtain $\partial\rho/\partial t$ from Eq.~(\ref{cn2}) and ignore the $O(\epsilon^3)$ terms. Then we substitute the result
into the spatial components of Eq.~(\ref{Div}). Therefore Euler's equation in the weak-field limit reads
\begin{equation}
\epsilon^2(\vec{v}\cdot\nabla)\vec{v}+\epsilon\left(\frac{\partial\vec{v}}{\partial t} +\nabla\Phi+\frac{\nabla p}{\rho}\right)+\alpha c^4\nabla\rho=0.
\label{euler}
\end{equation} 

This equation can be written in terms of the effective quantities to. The result is
\begin{equation}
\frac{\partial\vec{v}}{\partial t}+(\vec{v}\cdot\nabla)\vec{v}+\nabla\Phi+\frac{\nabla p_{\rm{eff}}}{\rho_{\rm{eff}}}=0.
\end{equation}
Now, we have a complete set of differential equations in the weak field limit governing the dynamics of a self-gravitating fluid. Using Eqs.~(\ref{cn2}), (\ref{euler}), the Poisson's Eq.~(\ref{n1}) (or \ref{poisson}), and also an equation of state, one can investigate the gravitational stability of a self gravitating fluid in the context of the $f(R,Q)$ gravity. 

\section{Jeans analysis in the EMSG }\label{jeans_EMSG}
Let us focus on a static, infinite, homogeneous, spherically symmetric fluid in the context of the EMSG. The question is when such a system can be locally fragmented under its own gravity? To find the answer, one should find the dispersion relation by linearizing the Poisson's Eq.~(\ref{n1}) or (\ref{poisson}) for the cases $f_{RR}= 0$ and $f_{RR}\ne 0$ respectively, and also the hydrodynamics Eqs.~(\ref{cn2}) and (\ref{euler}). 
The physical quantities are considered to be as $\mathsf{X}=\mathsf{X}_0+\mathsf{X}_1$, where $\mathsf{X}_1\ll\mathsf{X}_0$ and the ``0" (1) index indicates the background (perturbed) quantities.
For a static background system we have $\vec{v}_0=0$. Moreover, homogeneity implies that, $\rho_0$ and $p_0$ are constant. Also, we set the gravitational potential of the background to be constant. However these assumptions do not satisfy the background equations. Therefore, to avoid the underlying ambiguity, one may assume that the Poisson's equation can describe only the perturbed system. This assumption is known as the \textit{Jeans swindle} and is widely used even in the standard Newtonian (\cite{Binney}), and post-Newtonian (\cite{kazemi}) cases.
The resulting first order equations can be easily found as follows
\begin{eqnarray}
&& \frac{\partial\rho_1}{\partial t}+\rho_0\nabla\cdot\vec{v}_1=0,\label{fe}\\
&&\frac{\partial\vec{v}_1}{\partial t}+\nabla\Phi_1+\frac{\nabla p_1}{\rho_0}+\alpha c^4\nabla\rho_1=0,\label{se}\\
&&\nabla^2\Phi_1=\frac{\gamma  c^4}{2f_R}\left(\rho_1+4\alpha c^2\rho_0\rho_1 \right),\label{te} 
\end{eqnarray}
when $f_{RR}\neq 0$ then the last equation should be replaced by
\begin{eqnarray}
&&\nabla^2\Phi_1=\frac{\gamma c^4}{6f_R}\bigg[
4\rho_1+10\alpha c^2\rho_0\rho_1+\mathcal{H}
\bigg]\,,
\end{eqnarray}
where $\mathcal{H}$ is defined as 
\begin{equation}
\mathcal{H}=-\frac{\mathcal{M}^2}{4\pi}\int\frac{e^{-\mathcal{M}|\vec{r-r'}|}}{|\vec{r-r'}|}\left( \rho_1(\vec{r'})-2\alpha c^2\rho_0\rho_1(\vec{r'})\right) d^3r'\,.
\end{equation}
Now, taking the temporal derivative of Eq.~(\ref{fe}) and the divergence of Eq.~(\ref{se}), and also using the Poisson's Eq.~(\ref{te}), one can easily find the following result for the case of $f_{RR}=0$
\begin{eqnarray}
\nonumber&&\frac{1}{\rho_0}\frac{\partial^2\rho_1}{\partial t^2}-\frac{\gamma c^4(1+4\alpha c^2\rho_0)}{2f_R}\rho_1-\left(\frac{c_s^2}{\rho_0}+\alpha c^4 \right) \nabla^2\rho_1=0\,.\\
\end{eqnarray}
Similarly for the case of $f_{RR}\neq 0$ one may simply find
\begin{eqnarray}
\nonumber&&\frac{1}{\rho_0}\frac{\partial^2\rho_1}{\partial t^2}-
\frac{\gamma c^4}{6f_R}\bigg[
4\rho_1+10\alpha c^2\rho_0\rho_1+\mathcal{H}
\bigg]\\
&&~~~~
-\left(\frac{c_s^2}{\rho_0}+\alpha c^4 \right) \nabla^2\rho_1=0\label{a1}\,.
\end{eqnarray}
In the spherical coordinates system $(r,\theta,\varphi)$, using the Fourier form for the first-order perturbations
as $\rho_1=\rho_a\exp\left(i\left(\vec{k}\cdot\vec{r}-\omega t \right)  \right)$, one can simplify the dispersion relations. Therefore, Eq.~(\ref{a1}) can be straightforwardly integrated setting $\vec{r-r'}=\vec{R}$.  
Without loss of generality, one can assume that the wavenumber $k$ is along with the $z$ axis. So, after some manipulations it can be seen 
\begin{eqnarray}
\nonumber\mathcal{H}= && -\frac{\mathcal{M}^2}{4\pi}(1-2\alpha c^2\rho_0)\rho_1
\int\limits_{0}^{\infty}\int\limits_{0}^{\pi}\int\limits_{0}^{2\pi}\frac{e^{-\mathcal{M}R}}{R}e^{-i k R\cos\theta}\\
&&\times R^2\sin\theta dR d\theta d\varphi = -\frac{(1-2\alpha c^2\rho_0)}{1+\frac{k^2}{\mathcal{M}^2}}\rho_1\,.
\label{40}
\end{eqnarray}
Finally, the dispersion relation can be found as 
\begin{eqnarray}
\nonumber\omega^2 &&- (c_s^2+\alpha c^4\rho_0)k^2+ \frac{\gamma c^4\rho_0}{2 f_R}\\
&&\times\begin{cases}
(1+4\alpha c^2\rho_0)=0~~,~~f_{RR}=0\\
\\
\frac{1}{3}\left(4 +10\alpha c^2\rho_0-\frac{(1-2\alpha c^2\rho_0)\mathcal{M}^2}{\mathcal{M}^2+k^2} \right) =0~~,~~f_{RR}\ne0
\end{cases}
\label{41}
\end{eqnarray}
Now, let us investigate these two different cases with more detail.

\subsection{The case $f_{RR}$=0}\label{5_1}
%

Let us introduce two quantities that encode the modifications of the  dispersion relations :
\begin{equation}
\mathcal{C}_s^2=c_s^2+\alpha c^4\rho_0\label{Cs2}\,,
\end{equation}
\begin{equation}
\mathcal{G}=\frac{G}{f_R}(1+4\alpha c^2\rho_0)\label{geff}\,.
\end{equation}
Thus, the first expression in the dispersion relation (\ref{41}) can be recast as 
\begin{equation}
\omega^2-\mathcal{C}_s^2 k^2+4\pi \mathcal{G}\rho_0=0\,.\label{DRJ}
\end{equation}
This equation is similar to its Newtonian counterpart. In other words, turning the EMSG correction terms off, the Newtonian dispersion relation will be reproduced. From Eqs.~(\ref{Cs2})-(\ref{DRJ}) it is clear that EMSG for $f_R=1$ and $\alpha>0$ ($\alpha<0$) increases (decreases) both the sound speed and the gravitational strength effectively. However, these quantities have completely opposite impacts on the stability of the system. Therefore at the first sight it is not trivial to argue the final outcome of EMSG's corrections on the stability of the system. Nevertheless, it is important to mention that in our perturbative analysis we assumed that $|\alpha c^2\rho|\ll1$. Therefore, it is clear from Eq.~(\ref{geff}) that the EMSG effects do not change the effective gravitational strength significantly. On the other hand, the effective sound speed can be influenced substantially. One should note that we are working in a regime where the characteristic energy scale is high enough to allow EMSG effects to play role. In such a situation, since the sound speed $c_s^2$ is not necessarily much smaller than $c$, the correction term in (\ref{Cs2}) is not negligible. In other words, the ratio of two terms in the right hand side of (\ref{Cs2}), i.e., $|\alpha \rho c^2|\frac{c^2}{c_s^2}$ , should not be considered as a very small ratio.  Consequently, if we consider the sound speed as the representative of pressure content of the system, then one may accordingly infer that EMSG influences the effective pressure in the system. Now it make sense to conclude from (\ref{Cs2}) that if $\alpha>0$ ($\alpha<0$) then EMSG stabilizes (destabilizes) the fluid. In the following we compute a new Jeans wavenumber to clarify this issue.

By setting $\omega=0$ in Eq.~(\ref{DRJ}), we can obtain the border of stability. In this case, the new Jeans wavenumber in the context of EMSG can be obtained as
\begin{equation}
k_{JE}^2=\frac{4\pi \mathcal{G}\rho_0}{f_R \mathcal{C}_s^2}=k_J^2f_R^{-1}\Big(\frac{1+4\alpha c^2\rho_0}{1+\frac{\alpha c^4 \rho_0}{c_s^2}}\Big)\label{mah1}
\end{equation}
{
	where $k_J^2=4\pi G\rho_0/c_s^2$ is the standard Jeans wavenumber.
	Now, we can recast the dispersion relation in Eq.~(\ref{DRJ}) in a more useful form by introducing the following variables:
	\begin{eqnarray}
	&& \label{omegatilde} \tilde{\omega}= \frac{\omega}{\sqrt{4\pi G\rho_0}}\,,\\
	&& \label{kappatilde} \tilde{k}= \frac{k}{k_J}\,.
	\end{eqnarray}
	Thus, after some calculations, we obtain
	\begin{equation}
	\tilde{\omega}^2 - \left( 1+ \alpha\rho_0\frac{c^4}{c_s^2}\right) \tilde{k}^2 + \frac{1+4\alpha c^2\rho_0}{f_R}=0\,.
	\end{equation}
	Let us note that, once again, turning off the EMSG's correction terms the Newtonian case is obtained.

	Also, the standard Jeans wavelength, and Jeans mass can be introduced as $\lambda_J=2\pi/k_J$, and $\mathfrak{M}_J=\frac{4\pi}{3}\rho_0\left(\frac{\lambda_J}{2}\right)^3$ respectively, where $\mathfrak{M}_J$ is defined as the mass inside a sphere with radius $\lambda_J/2$. It can be shown that, the modified versions of the Jeans wavelength and mass respectively are
	
	\begin{equation}
	\lambda_{JE}^2=\lambda_J^2\,\Big(\frac{ 1+\alpha\rho_0 \frac{c^4}{c_s^2}}{1+4\alpha c^2 \rho_0}\Big)f_R\,.
	\end{equation}
	and
	\begin{equation}
	\mathfrak{M}_{JE}=\mathfrak{M}_{J} \left(\frac{\lambda_{JE}}{\lambda_J}\right)^3=\mathfrak{M}_{J}f_R^{3/2}\Big(\frac{ 1+\alpha\rho_0 \frac{c^4}{c_s^2}}{1+4\alpha c^2 \rho_0}\Big)^{\frac{3}{2}} \,.
	\label{mj}
	\end{equation}

	The fluid system can be more unstable (stable) in the context of EMSG than the Newtonian case, whenever $\mathfrak{M}_{JE}<\mathfrak{M}_{J}$ ($\mathfrak{M}_{JE}>\mathfrak{M}_{J}$). Eq.~(\ref{mj}) shows that, deviations from the standard case directly depends on the sign and value of the free parameter $\alpha$. It is obvious that,  negative (positive) values of $\alpha$ make the system more unstable (stable) in the context of EMSG with respect to the Newtonian case. Another less interesting point is that higher values for $f_R$ leads to higher Jeans masses. In other words, by increasing $f_R$ the system is stabilized. This is expected since $f_R$ reduces the effective gravitational constant, i.e., $G_{eff}\propto G/f_R$, and consequently weakens the destabilizing behaviour of gravitational force. However, we know that this parameter cannot deviate from unity significantly.

	Before moving on to close this subsection it is interesting to mention that in the original EMSG model \cite{us}, $\alpha=-\eta<0$. On the other hand this model leads to bouncing cosmological solutions and prevents the big bang singularity. As we showed, negative $\alpha$ destabilizes the local perturbations and supports the local gravitational  collapse. This behaviour seems completely in disagreement with the "stabilizing" behaviour of the theory in the early universe. However, one should note that here we present a non-relativistic description, while in the early universe we deal with a completely relativistic situation. 

	\subsection{The case $f_{RR}\neq 0:$}
For sake of completeness, we also compute the Jeans scale for the more general case $f_{RR}\neq 0$. 
In this case, we can recast {the second relation in the dispersion relation} in Eq.~(\ref{41}) in term of the variable in Eqs.~(\ref{omegatilde}) and (\ref{kappatilde}): 
\begin{equation}
\tilde{\omega}^2 + \frac{1}{3f_R}\biggl[ (4+10\alpha c^2\rho_0) - \frac{1-2\alpha c^2\rho_0}{1+\tilde{k}^2k_J^2\mathcal{M}^{-2}}\biggr]-\biggl[1+\frac{\alpha c^4\rho_0}{c_s^2}\biggr]\tilde{k}^2=0\,.
\end{equation}
Then, one can simply find the modified version of the Jeans wavenumber by setting $\tilde\omega^2=0$ {in this equation, and solve for $\tilde{k}^2$. In this case we found two solutions. One of these solutions recovers the standard Jeans wavenumber. This solution, without any expansion with respect to $\alpha$, reads}
\begin{eqnarray}
\nonumber && k_{JE}^2   =\frac{1}{6 \mathcal{C}_s^2 f_R} \biggl[2 c_s^2 k_J^2 (5 \alpha c^2  \rho_0+2)-3 \mathcal{C}_s^2 f_R \mathcal{M}^2+\\
&&+
\sqrt{(c_s^2 \mathcal{K}_1 + 3 f_R \mathcal{M}^2 \alpha c^2 \rho_0)^2 + 
	(6 c_s k_J\mathcal{M} \mathcal{C}_s)^2 f_R  \mathcal{K}_2}
\biggr]   
\label{mah2} 
\end{eqnarray}
where we have defined
\begin{eqnarray}
&&\mathcal{K}_1=3 f_R \mathcal{M}^2-2 k_J^2 (5 \alpha  c^2\rho_0+2)\,,\\
&&\mathcal{K}_2=\frac{\mathcal{G}f_R}{G}\,,   
\end{eqnarray}
and  we have chosen the solution which reproduces the standard Jeans wavenumber in the limiting case $\mathcal{M}\rightarrow\infty$ or equivalently $f_{RR}\rightarrow 0$. It is worth mentioning that the limit $\mathcal{M}\rightarrow 0$ does not recover the Newtonian results.

As our final remark in this section, one should take this case, i.e., $f_{RR}\neq 0$, with more care. As already mentioned, our analysis in this paper can be considered as a modification to the so-called metric $f(R)$ gravity theory. In order to make $f(R)$ gravity suitable for explaining the cosmic speed up, it seems necessary to include a very small scalar mass $\mathcal{M}$. Otherwise the extra scalar degree of freedom intrinsic in $f(R)$ gravity is not light enough to propagate in cosmic scales. Therefore it will not be effective for explaining the late time cosmic acceleration. However it is well-known in the relevant literature that if we take small $\mathcal{M}$, then the theory will have serious problems in the weak field limit and cannot recover the Newtonian gravity , for a review see \cite{PhysRept}. This is exactly what we see in our calculations for the Jeans wavenumber. In other words, we see that at the limit $\mathcal{M}\rightarrow 0$ Eq.~(\ref{mah2}) does not recover the Newtonian Jeans wavenumber. 

To address the above mentioned problem, it is necessary to take into account screening behaviour of $f(R)$ gravity theory \cite{hu2007models}. Investigating stability issues in the presence of screening effects, can be considered as a separate study. In this paper we continue our analysis without including the screening effects in the calculations for $f_{RR}\neq 0$, and put emphasis on the $f_{RR}=0$ case which does not suffer from the above mentioned problem. Therefore, hereafter we only discuss the  $f_{RR}=0$.
	\section{Toomre's criterion in EMSG}\label{tomre_emsg}
So far we have studied the stability of an infinite homogeneous medium without rotation. Nevertheless, the stability of rotating systems in EMSG is interesting in the sense that in high energy systems like HMNS, where we expect that EMSG contributions to be significant, the differential rotation is one of the main ingredients of the system. For simplicity, we restrict ourselves to rotating thin disks. On the other hand, for such a system there is already a well-known stability criterion in the standard Newtonian description known as Toomre's stability criterion \cite{Toomre1964}. In this case, one may simply compare the stability criteria based on EMSG with the Newtonian one.

To study the gravitational stability of a thin self-gravitating fluid disk, one should find the dispersion relation of propagating perturbations. This task can be addressed by particularizing the hydrodynamics equations of EMSG, given in Sec. \ref{Sect:hydroeq}, for a thin disk. One may conveniently assume that $\rho=\Sigma\delta(z)$, where $\Sigma$ is the surface density and $\delta$ is the Dirac's Delta function. Then the continuity Eq.~(\ref{cn2}) takes the following form
\begin{equation}
\frac{\partial\Sigma}{\partial t} +\nabla\cdot (\Sigma \vec{v})= - \frac{\alpha \Sigma \delta(z) c^2}{1+\alpha \Sigma \delta(z) c^2} \vec{v}\cdot \nabla \Sigma
\label{cn3}
\end{equation}
It is clear that the right-hand-side diverges on the plane of the disk.  This is not the case in Newtonian gravity. In fact, at $z=0$ we have $\alpha \Sigma \delta(z) c^2\rightarrow\infty$. This means that our linearized analysis based on the main assumption that $\alpha\rho c^2\ll 1$ is obviously violated. To skip this complexity and keep the analysis self-consistent, it is useful to assume a finite thickness for the disk. To do so, we simply assume that the density does not change in the vertical direction and is given by $\rho(r,\varphi)=\Sigma(r,\varphi)/l$ \cite{Toomre1964}, where $l$ is a small thickness of the disk and appears as a constant in our calculations. This method leads to a powerful estimation for the effect of the thickness on the stability of the disk \cite{1984ApJ...276..114J}. For $|z|> l/2$ the matter density vanishes $\rho=0$, and  the only constraint on the thickness is $l\gg \alpha \Sigma c^2$ everywhere on the disk. This condition guarantees the validity of our perturbative analysis. For a more careful way to include the thickness of the disk, we refer the reader to \cite{1970ApJ...161...87V} and \cite{1992MNRAS.256..307R}. Note that we are interested to the effects of EMSG and not the full analysis of the thickness. Therefore, it turns out that following the Toomre's method \cite{Toomre1964} is helpful here. Finally for completeness, we will shortly discuss the generalization of the other method, i.e., \cite{1970ApJ...161...87V}, in EMSG. 

Now for our disk with finite thickness, the continuity equation and the Euler's equation read
\begin{eqnarray}
&& \frac{\partial\Sigma}{\partial t} +\nabla\cdot (\Sigma \vec{v})\simeq - \frac{\alpha c^2 \Sigma}{l} \vec{v}\cdot \nabla \Sigma
\label{cn4}\,,\\
&& \frac{\partial\vec{v}}{\partial t}+(\vec{v}\cdot\nabla)\vec{v}+\nabla\Phi+\frac{\nabla p}{\Sigma}+\frac{\alpha c^4}{l}\nabla\Sigma=0\,.\label{Euler}
\end{eqnarray}
Note that in the Euler's equation the $p$ is a pressure defined as force per unit length. On the other hand the Poisson's equation for the case $f_{RR}=0$ is given by the following equation
\begin{equation}
\nabla^2\Phi=\frac{\gamma {c^4 \Sigma}}{2 l\,f_R}\left(1+\frac{2\alpha {c^2}}{l}\Sigma \right)\,.
\label{Poi2}
\end{equation}
In order to have a closed set of differential equations, we assumed the equation of state (EOS) to be barotropic,i.e., $p=p(\Sigma)$.
To study the stability of a disk we have to achieve the modified version of Toomre's criterion in the context of EMSG. To do so, one should linearize the hydrodynamics Eqs.~(\ref{cn4})-(\ref{Poi2}) in the cylindrical coordinate system $(r,\varphi,z)$. Eqs.~(\ref{cn4}) and (\ref{Euler}) in the cylindrical coordinate are
\begin{eqnarray}
\frac{\partial\Sigma}{\partial t}+\frac{1}{r}\frac{\partial}{\partial r}\left(\Sigma r v_{r}\right)+&&\frac{1}{r}\frac{\partial}{\partial\varphi}\left(\Sigma v_{\varphi}\right)+\frac{\partial}{\partial z}\left(\Sigma v_{z}\right)=\\&&- \frac{\alpha c^2 \Sigma}{l} \Big(v_r \frac{\partial \Sigma}{\partial r} + \frac{v_\varphi}{r} \frac{\partial \Sigma}{\partial \varphi}\Big)\,,
\label{cn5}
\end{eqnarray}
\begin{eqnarray}
\frac{\partial v_{r}}{\partial t}+v_{r} \frac{\partial v_{r}}{\partial r}+&&\frac{v_{\varphi}}{r} \frac{\partial v_{r}}{\partial \varphi}-\frac{v_{\varphi}^{2}}{r}+v_z \frac{\partial v_r}{\partial z}=\\&&- \frac{\partial}{\partial r}\left(\Phi+h\right)-\frac{\alpha c^4}{l}\frac{\partial \Sigma}{\partial r}\,,
\label{cn6}
\end{eqnarray}
\begin{eqnarray}
\frac{\partial v_{\varphi}}{\partial t}+v_{r} \frac{\partial v_{\varphi}}{\partial r}+&&\frac{v_{\varphi}}{r} \frac{\partial v_{\varphi}}{\partial \varphi}+\frac{v_{\varphi} v_{r}}{r}+v_z \frac{\partial v_{\varphi}}{\partial z}=\\&&- \frac{1}{r}\frac{\partial}{\partial\varphi}\left(\Phi+h\right)-\frac{\alpha c^4}{l r}\frac{\partial \Sigma}{\partial \varphi}\,,
\label{cn7}
\end{eqnarray}
\begin{eqnarray}
\frac{\partial v_{z}}{\partial t}+v_{r} \frac{\partial v_{z}}{\partial r}+&&\frac{v_{\varphi}}{r} \frac{\partial v_{z}}{\partial \varphi}+v_z \frac{\partial v_{z}}{\partial z}=\\&&-\frac{\partial}{\partial z}\left(\Phi+h\right)-\frac{\alpha c^4}{l }\frac{\partial \Sigma}{\partial z}\,,
\label{cn8}
\end{eqnarray}
for more details we refer the reader to \cite{Binney}.

Let us find the perturbed form of Eqs.~(\ref{cn5}) -(\ref{cn7}). We recall that, the physical quantities are considered to be as $\mathsf{X}=\mathsf{X}_0+\mathsf{X}_1$, where $\mathsf{X}_1\ll\mathsf{X}_0$ and the ``0" (1) index 
represents the background (perturbed) quantity. Moreover, the background is
assumed to be static and axis-symmetric, and the initial radial and vertical velocities vanish everywhere throughout the disk i.e., $v_{r0}=v_{z0}=0$. The only non-zero velocity component is $v_{\varphi0}$. We assume that $v_{\varphi0}$ is a function of radius and does not change in the $z$ direction. Of course, one can show that such barotropic equilibrium state with the density $\rho(r,\varphi)=\Sigma(r,\varphi)/l$ does not exists. Therefore, for the background system we are using a generalized version of the so-called \textit{Jeans-swindle}. Consequently although we do not care about the validity of the background system, the linear perturbations should satisfy all the linearized equations. The linearized versions of the continuity equation and also the radial, azimuthal and the vertical components of the Euler's equation read (the linearized Poisson's equation is discussed in the next subsection)
\begin{eqnarray}
\frac{\partial \Sigma_{1}}{\partial t}+&&\frac{1}{r}\frac{\partial}{\partial r}\left(\Sigma_{0} r v_{r1}\right)+\Omega \frac{\partial \Sigma_{1}}{\partial \varphi}+\frac{\Sigma_{0}}{r} \frac{\partial v_{\varphi 1}}{\partial \varphi}\\&&+\Sigma_0\frac{\partial v_{z1}}{\partial z}=-\frac{\alpha c^2\Sigma_0}{l}\Big(v_{r1} \frac{\partial \Sigma_0}{\partial r}+\Omega \frac{\partial \Sigma_1}{\partial \varphi}\Big)\,,
\label{continP}
\end{eqnarray}
\begin{equation}
\frac{\partial v_{r1}}{\partial t}+\Omega \frac{\partial v_{r1}}{\partial \varphi}-2\Omega v_{\varphi 1}=-\frac{\partial}{\partial r}\left(\Phi_{1}+h_{1}\right)-\frac{\alpha c^4}{l}\frac{\partial \Sigma_1}{\partial r}\,,
\label{EulerRP}
\end{equation}
\begin{equation}
\frac{\partial v_{\varphi 1}}{\partial t}+\Omega \frac{\partial v_{\varphi 1}}{\partial \varphi}+\frac{\kappa^{2}}{2\Omega} v_{r1}=-\frac{1}{r}\frac{\partial}{\partial \varphi}\left(\Phi_{1}+h_{1}\right)-\frac{\alpha c^4}{l\,r}\frac{\partial \Sigma_1}{\partial \varphi}\,,
\label{EulerPP}
\end{equation}
\begin{equation}
\frac{\partial v_{z 1}}{\partial t}+\Omega \frac{\partial v_{z 1}}{\partial \varphi}=-\frac{\partial}{\partial z}\left(\Phi_{1}+h_{1}\right)-\frac{\alpha c^4}{l}\frac{\partial \Sigma_1}{\partial z}\,,
\label{Eulerz}
\end{equation}
where $\Omega=v_{\varphi0}/r$ is the rotational frequency and $v_r$ and $v_{\varphi}$ are the radial and azimuthal components of velocity respectively.
Moreover, $\kappa=\sqrt{4\Omega^2+2r\Omega\Omega'}$ is the epicyclic frequency and also $h_1$ is defined as $h_1=c_s^2 \Sigma_1/\Sigma_0$. It should be noted that the prime stands for derivative with respect to $r$. It is worth mentioning that, by turning the correction terms (containing $\alpha$) off, the Newtonian hydrodynamics equations will be reproduced. Hereafter we define a new parameter $\alpha^*$ as
\begin{equation}
\alpha^*=\frac{\alpha}{l}\,.
\end{equation}

Now, by applying the WKB approximation one can {study the local stability of the disk} and {benefit} an elegant simplification as well. The general form of the perturbed quantities can be written as $\mathsf{X}_1=\mathsf{X}_a \exp\left(i\left(\vec{k}\cdot\vec{r}+m\varphi-\omega t \right)  \right) $, where $\omega$ is the oscillation frequency, $k$ is the wavenumber, and $m$ is a positive integer {which} determines the symmetry of the disturbances. For a tightly wound density wave, one can show that $|k r/m|\gg1$, therefore, using the WKB approximation the terms containing $1/r$ can be neglected comparing with the analogous terms containing $k$. This directly means that our description works only for local perturbations and one cannot use it for global stability of the disk. It is also necessary to mention that in order to recover the standard Toomre's criterion and regarding to the small thickness of the disk, we study perturbations propagating the $x-y$ plane. This means that the vertical component of $\mathbf{k}$ is zero. Finally, using Eq.~(\ref{continP}), the continuity equation can be written as follows 
\begin{equation}
[\omega-m \Omega (1+\alpha^* c^2 \Sigma_0)]\Sigma_a-k \Sigma_0 v_{ra}+i \Sigma_0 \frac{\partial v_{az}}{\partial z}=0\,.
\label{continF}
\end{equation}

Moreover, considering Eqs. (\ref{EulerRP}) and (\ref{EulerPP}), the solutions for coefficients of Fourier expansions of perturbed velocity components can be found as below
\begin{equation}
v_{r a}=\frac{(m\Omega-\omega)k}{\Delta}\left(\Phi_a+h_a+\alpha^*{c^4}\Sigma_a \right),\label{EulerRF}
\end{equation}
\begin{equation}
v_{\varphi a}=-\frac{2B i k}{\Delta}\left(\Phi_a+h_a+\alpha^*{c^4}\Sigma_a \right),\label{EulerPF}
\end{equation}
\begin{equation}
v_{za}=\frac{i}{m\Omega-\omega}\frac{\partial \Phi_a}{\partial z},\label{EulerPZ}
\end{equation}
where rotation Oort's constant $B$ and $\Delta$ are defined as below
\begin{eqnarray}
&& B(r)=-\frac{1}{2}\left(\Omega+\frac{d(\Omega r)}{dr} \right)\,,\\
&& \Delta=\kappa^2-(m\Omega-\omega)^2\,.
\end{eqnarray}
In order to find the dispersion relation, the next step is to find the potential of a WKB spiral pattern $\Sigma_1=\Sigma_a\exp\left(i(\vec{k}\cdot\vec{r}+m\varphi-\omega t) \right) $ to determine $\Phi_a$ in terms of $\Sigma_a$.

\subsection{The gravitational potential of a thick WKB density wave in Newtonian gravity and EMSG}
Now let us calculate the gravitational potential of a WKB density wave in the context of EMSG. The linearized version of the modified Poisson's Eq.~(\ref{Poi2}) is
\begin{equation}
\nabla^2 \Phi_1=\frac{\gamma c^4}{2 l\, f_R}(1+4 \alpha^* c^2 \Sigma_0)\Sigma_1=\frac{4\pi \mathcal{G}}{l}\Sigma_1
\label{poi5}
\end{equation}
where we have used the definition $\mathcal{G}=G/f_R (1+4\alpha^* c^2 \Sigma_0)$ introduced in (\ref{geff}) and replaced $\alpha$ with $\alpha^*$. Let us note that all the calculations in this subsection also hold in Newtonian gravity, we just need to set $\mathcal{G}\rightarrow G$. We know that the density wave $\Sigma_1$ in the WKB approximation at arbitrary location $\mathbf{r}$ on the disk, can be considered a plane wave propagating in the radial direction. Therefore, without loosing of generality we take $\vec{k}$ along $\hat{\vec{x}}$. So finding the potential of a WKB wave reduces to finding the potential of a plane density wave in EMSG. Consequently, one may write  $\Sigma_1=\Sigma_a \exp i( k x- \omega t)$. For this plane wave we guess the potential has the following functional form
\begin{equation}
\Phi_1(x,z,t)= \Phi_a e^{ i( k x- \omega t) }f(z)\,.
\label{poi6}
\end{equation}
Substituting the above solution into Eq.~(\ref{poi5}), we get the following differential equation for the function $f(z)$
\begin{equation}
\frac{d^2 f}{d z^2}-k^2 f(z)=\frac{4\pi \mathcal{G}}{l} \frac{\Sigma_a}{\Phi_a}\,.
\label{e1}
\end{equation}
This equation holds for $|z|<l/2$. On the other hand we know that for $|z|>l/2$ the potential is given by
\begin{equation}
\Phi_{out}=\Phi_a e^{i (k x -\omega t)} e^{- k |z|}\,.
\end{equation}
Note that hereafter we restrict ourselves to trailing density waves with $k>0$. Eq.~(\ref{e1}) can be simply integrated to obtain $f(z)$, and the integration constants will be fixed using the following matching conditions
\begin{equation}
\Phi_1(z=\pm l/2)=\Phi_{out}(z=\pm l/2)\,.
\end{equation}
Thus, we obtain
\begin{eqnarray}
\Phi_a f(z)=&&\frac{ \text{cosh}(k z) \text{sech}(k l/2)-1}{k^2}\frac{4\pi \mathcal{G}}{l}\Sigma_a\\&&
+(1-\text{tanh}(k l/2))\text{cosh}(k z)\Phi_a\,.
\label{poi10}
\end{eqnarray}
On the other hand, we expect that at the plane of the disk ($z=0$) and in the limit $l\rightarrow 0$, the standard thin disk potential should be recovered. Therefore let us choose $\Phi_a$ as follows
\begin{equation}
\Phi_a=-\frac{\text{sinh}\frac{kl}{2}}{\frac{kl}{2}}\frac{2 \pi \mathcal{G}}{k}\Sigma_a\,,
\label{poi20}
\end{equation}
it is clear that at the limit $l\rightarrow 0$ the standard thin disk density wave, $\Phi_a=-\frac{2 \pi \mathcal{G}}{k}\Sigma_a$ \cite{Binney}, is recovered. Combining Eqs.~(\ref{poi10}) and (\ref{poi20}) we find the final form of the potential as
\begin{equation}
\Phi_1=-\frac{ e^{-\frac{1}{2} k (l+2 z)}
	\left(2 e^{\frac{k l}{2}+k z}-e^{2 k z}-1\right)}{k l}\frac{2 \pi  \mathcal{G} }{k}\Sigma_1\,,
\end{equation}
and   at $z=0$ we have
\begin{equation}
\Phi_1=-\frac{1-e^{-\frac{k
			l}{2}}}{k l/2}\frac{2 \pi  \mathcal{G}}{k}\Sigma_1=-\mathcal{F}(kl)\frac{2 \pi  \mathcal{G}}{k}\Sigma_1\,.
\label{poi21}
\end{equation}
The reduction factor $\mathcal{F}(kl)=\frac{ 1-\exp{(-kl/2})}{k l/2}$ is exactly the coefficient derived in \cite{Toomre1964}. This reduction coefficient can be interpreted as a decrease in the surface density. Consequently, it is well-established in the literature that thickness of the disk has stabilizing effects. 

\subsection{The dispersion relation and the Toomre's parameter}\label{sec6_2}
It is straightforward to show that the vertical average value of $\frac{\partial \Phi_1}{\partial z}$, namely its integration over $(-l/2,l/2)$, vanishes. Therefore, we use the approximation $\frac{\partial \Phi_1}{\partial z}\simeq 0$. Consequently, we have $v_{za}\simeq 0$. This assumption is another reason for taking Toomre's method as an estimation and not a precise calculation. Now, substituting Eqs.~(\ref{poi21}) and (\ref{EulerRF}) into Eq.~(\ref{continF}), and confining ourselves to the plane $z=0$, we find the following dispersion relation for axisymmetric ($m=0$) density waves
\begin{equation}
\omega^2=\kappa^2+\mathcal{C}_s^2 k^2-\mathcal{F}(k l)2\pi \mathcal{G}\Sigma_0 k\,,
\label{dis1}
\end{equation}
where the effective sound speed $\mathcal{C}_s^2=c_s^2+\alpha^* c^4\Sigma_0$ is defined in (\ref{Cs2}), again replacing $\alpha$ with $\alpha^*$.
At the limit $\alpha^*\rightarrow 0$,  we have $\mathcal{C}_s=c_s$ and $\mathcal{G}=G$. Therefore, as expected, the dispersion relation (\ref{dis1}) reduces to the standard one in Newtonian gravity \cite{1984ApJ...276..114J}. As we already mentioned, for $\alpha^*>0$, EMSG effects can be interpreted as an increase in the sound speed and in the gravitational constant as well. Increase in the sound speed, stabilizes the system while increase in the gravitational strength promotes the instability. Therefore, a careful analysis is required to discriminate between these opposite features. To do so, let us find the generalized version of the Toomre's criterion in EMSG.

Using the dispersion relation (\ref{dis1}), the stability condition $\omega^2>0$ takes the following form
\begin{equation}
\mathcal{Q}(X)^2>-\frac{4 X^2 \left(\beta + e^{-\frac{\beta }{X}}-1\right)}{\beta}\,,
\label{dis2}
\end{equation}
where the dimensionless wavelength $X$ is defined as $X=k_{\text{crit}}/k$ and $k_{\text{crit}}=\kappa^2/(2 \pi \mathcal{G}\Sigma_0)$. Furthermore the $\beta$ parameter as the representative of the thickness of the disk is defined as $\beta=k_{\text{crit}} l/2$. Before discussing the effects of EMSG, let us briefly review the impact of thickness on the stability of the disk. Our discussion here is true in both Newtonian gravity and EMSG. It is straightforward to verify that for $\beta\ge 1$, the right hand side of (\ref{dis2}) gets negative. This means that all the perturbations would be stable. Note that for large $\beta$, the characteristic length of the system in the vertical direction increases. Therefore, one may expect the ordinary Jeans's criterion accounts for the stability of the system. At this limit Eq.~(\ref{dis1}) is written as
\begin{equation}
\omega^2\simeq\mathcal{C}_s^2 k^2-4 \pi \mathcal{G}\rho+\kappa^2\,,
\label{dis22}
\end{equation}
if we ignore the angular momentum in the system, i.e., $\kappa=0$, then the well-known dispersion relation already derived in Jeans analysis of an infinite medium is recovered, see Eq.~(\ref{DRJ}). If the combination of the last two terms in the right hand side of (\ref{dis22}) gets positive, or equivalently if $\beta\ge 1$, the all the wavelengths will be stable.

The other more interesting case is $\beta\le 1$. In this case we directly use the dispersion relation (\ref{dis2}) to find the stability criterion for each wavelength $X$. The boundary of stability, namely the minimum value required for $\mathcal{Q}(X)$ to stabilize the wavelength $X$ is shown in Fig.~\ref{border}.   In this figure, darker colors show larger values of $\beta$. Moreover, the inner surface of each curve supposed to be the unstable area. Therefore, it is clear that, the larger values of $\beta$ decrease the instability area. In other words,
this figure directly shows that increasing the $\beta$ parameter, increases the stability of the disk. As we mentioned, from this perspective, both Newtonian and EMSG behave in a similar way. 

So far we considered the stabilizing effects of the disk thickness. Now, let us investigate our main purpose in this section: impact of EMSG on the stability of self-gravitating disks. To do so, one should note that $\mathcal{Q}$ in (\ref{dis2}) has been written in terms of effective parameters $\mathcal{C}_s$ and $\mathcal{G}$. Furthermore the epicycle frequency $\kappa$ is different from Newtonian case in the sense that it includes EMSG corrections. Therefore for comparison with Newtonian gravity, it is helpful to rewrite (\ref{dis2}) in terms of the Newtonian Toomre's parameter $Q=\kappa_N c_s/\pi G\Sigma_0$.  Where $\kappa_N$ is the epicycle frequency obtained using the Newtonian gravitational force. For a given matter density $\Sigma_0$, let us express the epicycle frequency as
\begin{equation}
\kappa^2=\kappa_N^2+\delta\kappa^2\,,
\end{equation}
where $\delta\kappa^2$ is the corrections to $\kappa_N^2$ induced by EMSG. We expect this correction be proportional to $\alpha^*$. Accordingly we have $X=X_N+\delta X$ and $\beta=\beta_N+\delta\beta$, where $\delta X$ and $\delta\beta$ are also proportional to $\alpha^*$. Now, we rewrite Eq.~(\ref{dis2}) as
\begin{equation}
Q(X_N)^2>-\frac{4 X_N^2 \left(\beta_N f_R +e^{-\frac{\beta_N }{X_N}}-1\right)}{\beta_N f_R}+\alpha^* \Delta+\mathcal{O}(\alpha^{*2})\,,
\label{dis4}
\end{equation}
where $\Delta$ is a complicated function of $\delta\beta$, $\delta\kappa$, $\delta X$, $\beta_N$, $X_N$ and $\kappa_N$. So we avoid to write it here. This term includes all the corrections introduced by EMSG. It is difficult to specify the sign of $\alpha^* \Delta$. We know that if $\alpha^* \Delta<0$ ($>0$) then EMSG stabilizes (destabilizes) the disk. Furthermore we need a known surface density $\Sigma_0$ to calculate all the functions in Eq.~(\ref{dis4}) and quantify the differences of EMSG and Newtonian gravity. In the next section we study an exponential toy model in order to  describe the EMSG impact on the stability of disks.
\begin{figure}
	\centering
	\includegraphics[scale=0.6]{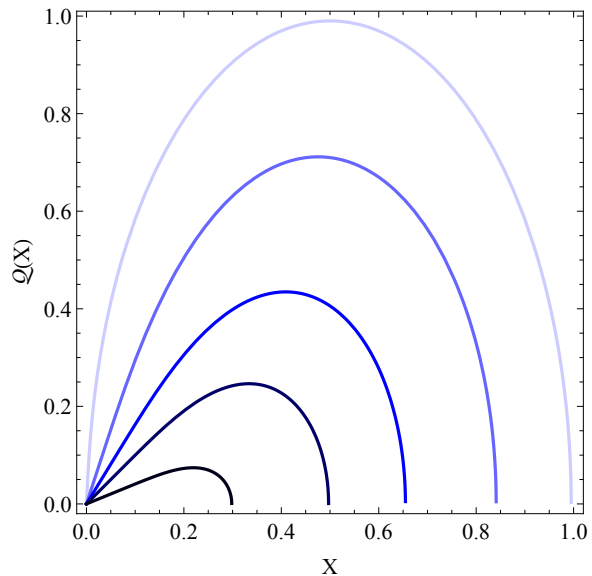}
	\caption{Curves from up to down belongs to $\beta= 0.01$, $0.3$, $0.6$, $0.8$, $0.96$. This shows that thickness of the disk seriously stabilizes the disk. Of course, one should note that we have used an estimative way in our analysis.}
	\label{border}
\end{figure}

Before closing this section it is worthy to mention that we used an estimative method to find the Toomre's criterion. Interestingly we found that the reduction factor $\mathcal{F}$ in the dispersion relation is the same as in Newtonian gravity. On the other hand, we know that a more precise method to include the thickness of the disk in Newtonian gravity leads to a different reduction factor as $\mathcal{F}=(1+ kl/2)^{-1}$ \cite{1970ApJ...161...87V}. Based on what happened in out estimative method, one may expect that the above mentioned reduction factor appears in EMSG as well. In this case the dispersion relation may be written as
\begin{equation}
\omega^2=\kappa^2+\mathcal{C}_s^2 k^2- 2\pi \mathcal{G}\Sigma_0\frac{k}{1+ k l/2}\,.
\end{equation}
However, we continue working with the estimative method explained comprehensively in this section.

	\section{Exponential fluid disk in the context of the EMSG}\label{disk_emsg}
Here, we are going to achieve a modified version of the Toomre's criterion using a common toy model. In fact, this model could help us to compare the results in the context of EMSG and Newtonian gravity. The EMSG effects appear in the frequency parameter $\kappa$ and effective parameters $\mathcal{C}_s$ and $\mathcal{G}$. Consequently, as mentioned earlier, it seems that, it is not straightforward to compare the new criterion  with the standard one. However, by specifying the surface density profile $\Sigma_0$, one can compare both theories. 

The exponential surface density profile is widely used to model wide variety of astrophysical systems. In this section we take the following exponential model as a toy model to clarify some differences between EMSG and GR in the weak field limit
\begin{equation}
\Sigma_0=\Sigma_p e^{-2y}.
\label{density_profile}
\end{equation}
Here, $y=r/2R_d$ is a dimensionless radius and $\Sigma_p$ and $R_d$ are the central density and a characteristic length scale respectively. 
Taking such a density profile, by solving the Poisson's equation in the Newtonian regime, one can show that the gravitational potential of a razor thin disk reads
\begin{equation}
\Phi_0(y,z=0)=-2\pi G \Sigma_p R_d y\left[ I_0(y)K_1(y)-I_1(y)K_0(y) \right]
\label{Npot} 
\end{equation}
where $I_n$ and $K_n$ for $n=0,1$, are modified Bessel functions of the first and second kinds, respectively (For more details see \cite{Binney}). We need to find the gravitational potential of a disk with a small thickness in the context of EMSG. It is clear that, Eq.~(\ref{Poi2}) can be written as
\begin{equation}
\nabla^2\Phi_0=\frac{4\pi G}{l f_R}\left(\Sigma_0+2\alpha^* c^2\Sigma_0^2 \right).\label{poi_Sigma}
\end{equation}
Note that, we will assume $f_R=1$ in the following. However, it can be recovered in the results by replacing  $G$ by $G/f_R$. 
It is clear from Eq.~(\ref{poi_Sigma}) that in order to find the potential in EMSG, we can simply add the Newtonian potentials of two separate disks with small thicknesses $l$, and mass densities $\Sigma_0/l$ and $2\alpha^* c^2\Sigma_0^2/l$. Therefore all we need is to find the gravitational potential of a thick disk with exponential functionality in the radial direction in Newtonian gravity. To do so, we find the gravitational potential of a thin disk along the $z$ axis and then integrate over thin disks to find the gravitational potential of a thick disk. Let us begin with finding the Newtonian gravitational potential of a razor thin and exponential disk, with the density profile given by Eq.~(\ref{density_profile}).
The gravitational potential for the field points that situated on the $z$ axis could be found as
\begin{eqnarray}
\nonumber\Phi_{\text{out}}(r=0,z) && =-G\int\frac{\Sigma_0 dA}{|\vec{r}-\vec{r}'|}\\
&& = -G\Sigma_p\int_{0}^{\infty}\int_{0}^{2\pi}\frac{e^{-r/R_d}r'dr'd\varphi}{\sqrt{r^{'2}+z^2}},
\end{eqnarray}
the integral can be simply solved, see \cite{2007tisp.book.....G}, to give
\begin{equation}
\Phi_{\text{out}}(y=0,z)=\pi ^2 G \Sigma_p \left| z\right| 
\left(\vec{H}_{-1}\left(\frac{\left| z\right|
}{R_d}\right)+Y_1\left(\frac{\left| z\right|
}{R_d}\right)\right),\label{phi_out}
\end{equation}
where, $\vec{H}$ and $Y$ denotes the Struve function and the Bessel function of the second kind respectively. 
Note that, both of these functions are well-behaved.
On the other hand, the gravitational potential on the surface of the disk plane is given by Eq.~(\ref{Npot}), i.e., $\Phi_{\text{in}}(y,z=0)=\Phi_0(y, z=0)$.
It is worth mentioning that, since the potential $\Phi$ is a continues function, it is easy to show that
\begin{equation}
\lim\limits_{y\rightarrow0}\Phi_{\text{in}}(y,z=0)=\lim\limits_{z\rightarrow 0}\Phi_{\text{out}}(y=0,z)=-2\pi G\Sigma_p R_d.
\label{phi_in}
\end{equation}
Now, the gravitational potential of a razor thin disk all over the space could be found combining Eqs.~(\ref{phi_out}) and (\ref{phi_in}) obtaining
\begin{eqnarray}\label{phi_phi}
\Phi_{\text{out}}(y=0,z) && =-2\pi  G \Sigma_p R_d\\\nonumber
&& \times \left(-\frac{\pi\left| z\right|}{2R_d} \right)  
\left(\vec{H}_{-1}\left(\frac{\left| z\right|
}{R_d}\right)+Y_1\left(\frac{\left| z\right|
}{R_d}\right)\right),
\end{eqnarray}
where, the first term at the right hand side, denotes $\Phi_{\text{in}}$ at $y\rightarrow 0$ limit. Therefore, regarding Eqs.~(\ref{phi_in}) and (\ref{phi_phi}) and keeping in mind that the gravitational potential can be separated in terms of vertical and radial coordinates as $\Phi(y,z)=f_1(z)f_2(y)$ , one can find the gravitational potential of a thin disk over the whole space as follows
\begin{eqnarray}\label{phi_1}
\nonumber\Phi(y,z) && =-2\pi G\Sigma_p R_d\left(y\left[ I_0(y)K_1(y)-I_1(y)K_0(y) \right] \right)\\
&&  \times\left(-\frac{\pi\left| z\right|}{2R_d} \right)  
\left(\vec{H}_{-1}\left(\frac{\left| z\right|
}{R_d}\right)+Y_1\left(\frac{\left| z\right|
}{R_d}\right)\right).
\end{eqnarray}
It can be shown that for the $z\rightarrow 0$ limit, the potential of Eq.~(\ref{Npot}) will be reproduced.

In the next step, we are going to describe the calculation of the gravitational potential of a thick disk. Here, the density profile of the thick disk assumed to be
\begin{equation}
\rho_0(y,z)=\Sigma_0(y)\zeta(z),
\end{equation}
where $\Sigma_0(r)$ is an exponential function as in (\ref{density_profile}) and $\zeta(z)=1/l$. We have chosen this special form for $\zeta(z)$ to be completely self-consistent with our calculations in the previous section. Of course one can use more realistic functions like $\zeta(z)\propto e^{-\mu z}$.

As already mentioned, a thick disk can be considered as a set of many infinitesimal thin layers with thicknesses $dz'$.   
The potential of each layer that situated at the  vertical distance $dz'$ from the disk, at the field point $(y,z)$ reads
\begin{equation}
d\tilde{\Phi}_0(y,z)=dz' \Phi(y,z-z')\zeta(z').
\end{equation}
By adding the contributions of all layers, the result will be
\begin{equation}\label{phi_2}
\tilde{\Phi}_0(y,z)=\int_{-\infty}^{\infty}dz' \Phi(y,z-z')\zeta(z').
\end{equation}
It should be noted that, since we are interested in the gravitational effects inside the disk, let us restrict ourselves to the equatorial plane $z=0$. So, using Eqs.~(\ref{phi_1}) and (\ref{phi_2}), one can see
\begin{eqnarray}
\tilde{\Phi}_0(y,z=0) = && \int_{-l/2}^{+l/2} \frac{dz'}{l}(-2\pi G\Sigma_p R_d)\\
&&\nonumber\times\left(y\left[ I_0(y)K_1(y)
-I_1(y)K_0(y) \right] \right)\\
&&
\nonumber\times \left(-\frac{\pi\left| z'\right|}{2R_d} \right)  
\left(\vec{H}_{-1}\left(\frac{\left| z'\right|
}{R_d}\right)+Y_1\left(\frac{\left| z'\right|
}{R_d}\right)\right).
\end{eqnarray}
Finally, after some manipulations, the integral can be analytically solved. The result reads
\begin{eqnarray}\label{phi_f}
&& \tilde{\Phi}_0(y)  =\frac{\pi  c^2 \eta  y (I_1(y) K_0(y)-I_0(y) K_1(y))
}{2 \xi }\\
\nonumber&& ~~\times\left(8 \pi  G_{1,3}^{2,0}\left(\frac{\xi ^2}{16}\bigg|
\begin{array}{c}
1 \\
\frac{1}{2},\frac{3}{2},0 \\
\end{array}
\right)-\xi ^2 \,
_2F_3\left(1,1;\frac{1}{2},\frac{3}{2},2;-\frac{\xi
	^2}{16}\right)\right),
\end{eqnarray}
where, the functions $G$ and $F$ that appears in this equation are the MeijerG and the  generalized hypergeometric functions respectively. Also, the dimensionless constants $\eta$ and $\xi$ are defined as
\begin{equation}
\eta=G R_d\Sigma_p/c^2 ~~~,~~~ \xi=l/R_d.
\end{equation}
It should be noted that, since the density falls off much faster along the $z$ axis than in the radial direction within the plane, the galactic disks assumed to be thin (\cite{Binney}). Although we have not restricted our analysis to galactic disks, we will keep the thin disk approximation in the subsequent sections.
So, the Eq.~(\ref{phi_f}) could be  expanded for the small values of $\xi$. By keeping only the linear terms of $\xi$, the result reads
\begin{equation}\label{phi_thick}
\tilde{\Phi}_0(y)  \simeq-\frac{1}{2} \pi  c^2 (\xi -4) \eta  y (I_1(y)
K_0(y)-I_0(y) K_1(y)).
\end{equation}
Again, as expected, it is clear that, turning off the contribution of thickness, the potential of a razor thin disk will be reproduced (see the Eq.~\ref{Npot}).

Now, to find the complete form of the gravitational potential in the context of EMSG, one can apply the following replacements to Eq.~(\ref{phi_thick}), 
\begin{equation}
y\rightarrow 2y ~~,~~  R_d\rightarrow \frac{R_d}{2} ~~,~~  \Sigma_p\rightarrow 2\alpha^* c^2\Sigma_p^2.
\end{equation}
The overall result is as follows 
	\begin{align}\label{Epot} 
\nonumber\tilde{\Phi}_0(y) & =\frac{ \pi  c^2 \eta  y}{2} \big(-4 \mathcal{A} (\xi -2)
	(I_1(2 y) K_0(2 y)\\\nonumber
	& -I_0(2 y) K_1(2 y))
	-(\xi -4)
	(I_1(y) K_0(y)\\
	&-I_0(y) K_1(y))\big)
	\end{align}
where $\mathcal{A}$ is a dimensionless parameter defined as
\begin{equation}
\mathcal{A}=\alpha^* \Sigma_p c^2.
\end{equation}
Hereafter we remove the tilde sign over the potential $\Phi_0$.

Now, as the next step, one can find the epicycle frequency $\kappa$. Using the radial component of  Eq.~(\ref{Euler}), i.e., Eq.~(\ref{cn6}), it can be shown that, the rotational frequency reads
\begin{equation}\label{Omega2}
\Omega^2=\frac{1}{4R_d^2 y}\left(\frac{\partial\Phi_0}{\partial y}+\frac{1}{\Sigma_0}\frac{\partial p_0}{\partial y}+\alpha^* c^4\frac{\partial\Sigma_0}{\partial y} \right).
\end{equation}
Using this equation one can find an analytic expression for the epicycle frequency in terms of radius. Now, following notation of the previous section, the epicycle frequency can be written as
\begin{equation}
\kappa^2=\kappa_N^2+\delta\kappa^2\label{kappa}
\end{equation}
where $\kappa_{\small N}^2$ is the Newtonian part of the epicycle frequency and $\delta\kappa^2$ is defined to parameterize the EMSG corrections. These functions are given by the following expressions

\begin{eqnarray}
\nonumber \kappa_N^2 && =\frac{c^2}{2 R_d^2} \bigg( \frac{\Gamma  \mu  e^{-2 (\Gamma -1) y} (2 (\Gamma -1)
	y-3)}{y}+\pi  \eta  (\xi -4) \\\nonumber
&&\times((y I_0(y)+I_1(y)) K_1(y)-(2 I_0(y)+y
I_1(y)) K_0(y))\bigg)\\
\label{kappa_N}
\end{eqnarray}
\begin{eqnarray}
\nonumber\delta\kappa^2= &&\frac{c^2\mathcal{A}}{2 R_d^2}\bigg(  8 \pi  \eta  (\xi -2) ((2 y I_0(2 y)+I_1(2 y)) K_1(2
y)\\\nonumber
&& -2 (I_0(2 y)+y I_1(2 y)) K_0(2 y))+\frac{e^{-2 y} (2
	y-3)}{y}\bigg)\\
\end{eqnarray}
The new dimensionless parameter $\mu$ is a representative of the sound speed and defined as follows
\begin{equation}
\mu=\frac{K\Sigma_p^{\Gamma-1}}{c^2}.
\end{equation}
In fact, the nature of this definition could be explored by picking an EOS. Here we have used the polytropic EOS 
\begin{equation}
p=K\Sigma_0^{\Gamma}\,,
\end{equation}
where $\Gamma$ is the polytropic index. Then, it is straightforward to show that the sound speed could be written as
\begin{equation}
c_s^2= K\Sigma_p^{\Gamma-1}\Gamma e^{-2y(\Gamma-1)}=c^2\mu\Gamma e^{-2y(\Gamma-1)}.
\label{css}
\end{equation}
Now we are in a position to define the modified version of the Toomre's criterion. 
	
\subsection{Toomre's criterion for an exponential disk in EMSG}

\begin{figure*}
	\centering
	\includegraphics[scale=0.59]{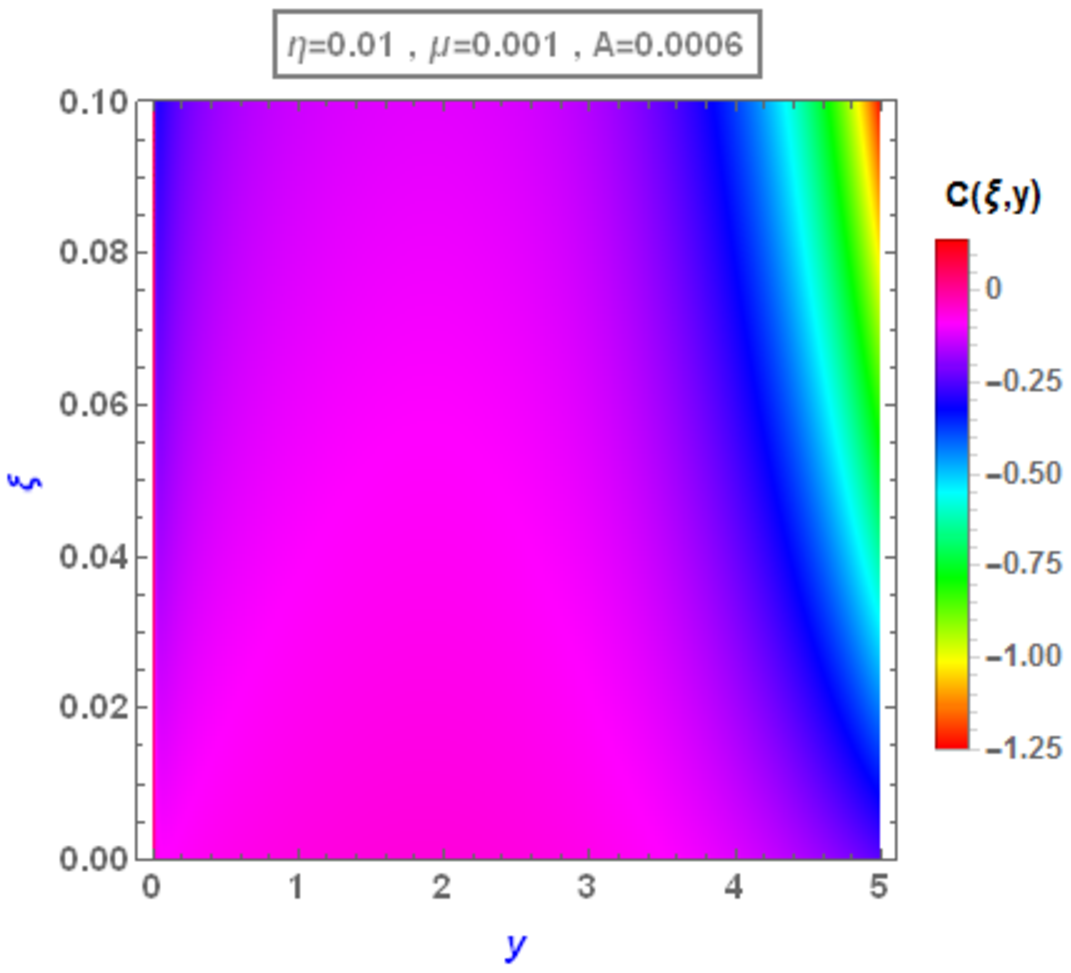}\hspace{0.0cm}
	\includegraphics[scale=0.62]{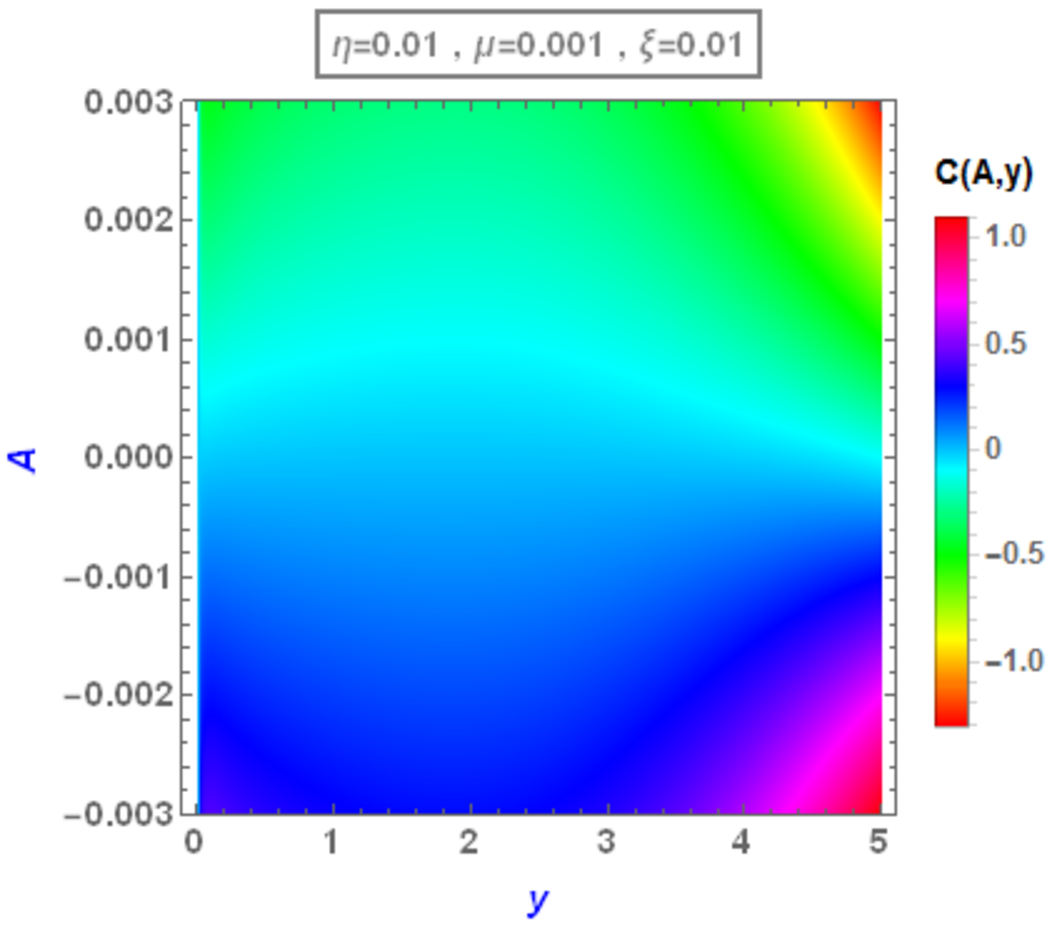}\vspace{0cm}
	\includegraphics[scale=0.6]{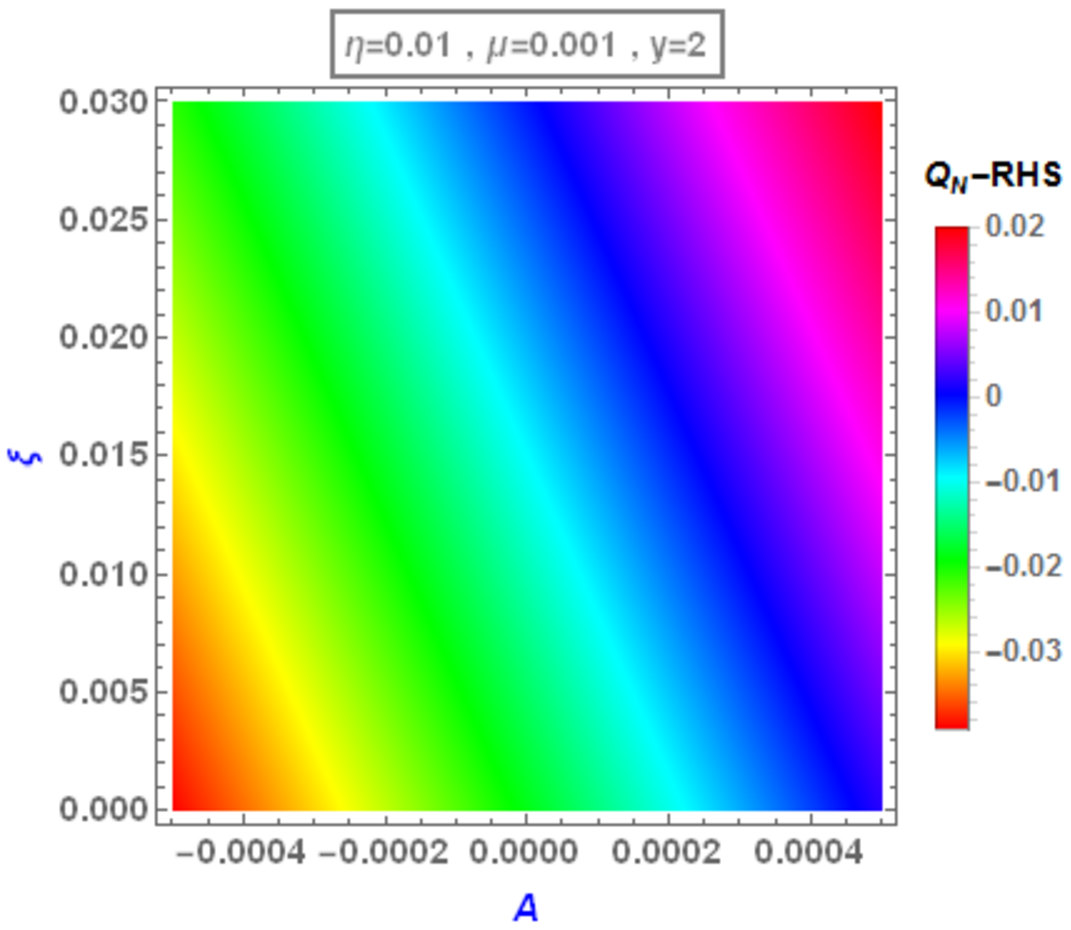}\hspace{0mm}\vspace{0cm}
	\includegraphics[scale=0.62]{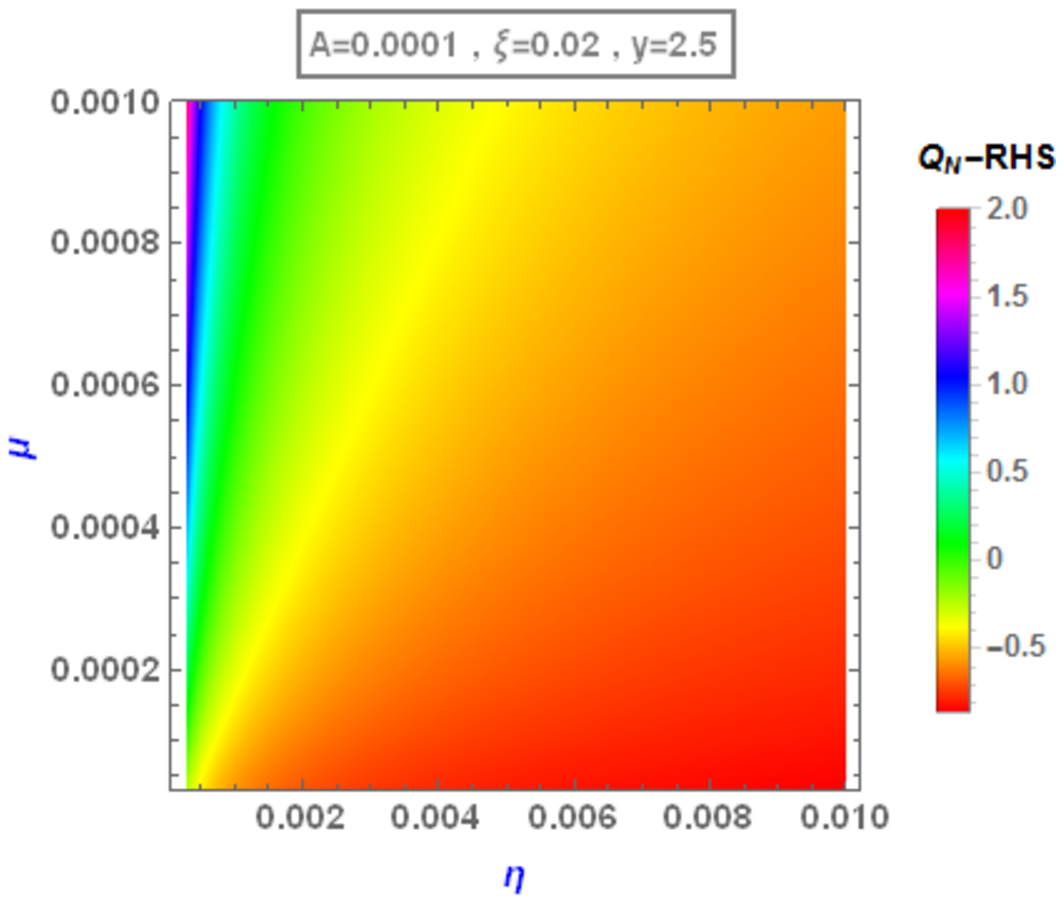}
	\caption{The correction term $\mathcal{C}$ and the Toomre's criterion in EMSG for various values of the model parameters. Note that, the Toomre's criterion of Eq.~(\ref{TC}) is written here as $\mathcal{Q}_N-RHS$. Whenever this expression is positive, the stability condition will be held.
		Furthermore, we take $\Gamma=2$.  \label{DPlots}}
\end{figure*}

First, we want to emphasis that the main aim of this paper is to study the role of  EMSG. On the other hand, the effect of thickness has been studied in Sec.~\ref{sec6_2} in order to overcome some technical difficulties. So, assuming $l=\xi R_d$, where $\xi\ll 1$, one can easily expand the dispersion relation of Eq.~(\ref{dis1}) up to $O(\xi)$. 
Note that,  although this simplification may not include all the real physical properties, it can provide a way to track the footprints of the EMSG effects.
Considering this point, the Eq.~(\ref{dis1} could be written as following
\begin{equation}
\omega^2=\kappa^2+\left(\mathcal{C}_s^2+\frac{\pi\mathcal{G}R_d\xi\Sigma_0}{2} \right)  k^2-2\pi \mathcal{G}\Sigma_0 k+O(\xi^2).
\label{disf}
\end{equation}
One can see that, replacing the effective quantities $\mathcal{C}_s$, and $\mathcal{G}$, with the Newtonian values, and also ignoring the thickness $l$ (or equivalently setting $\xi=0$), the Newtonian dispersion relation will be reproduced. As we already mentioned, disk thickness has stabilizing effects. This fact is clearly seen in the dispersion relation (\ref{disf}where the thickness parameter $\xi$ appears with a positive sign on the right hand side. Of course one should take this description more carefully in the sense that thickness also appears in $\kappa$.
On the right hand side (hereafter RHS) of Eq.~(\ref{disf}) all the quantities are real, therefore $\omega^2$ will be real too. So, if $\omega^2>0$, then $\omega$ is real and the disk is stable. On the other hand, there will be some growing modes, if $\omega^2<0$ and therefore, the disk will be unstable. 
Since the RHS of Eq.~(\ref{disf}) is a quadratic function of $k$, and also, the coefficient of $k^2$ is positive (note that the correction terms assumed to be smaller than the main terms), one can seek for a $k$ at which the RHS is minimum. This wavenumber reads
\begin{equation}\label{kmin}
k_{\text{min}}=\frac{\pi  \mathcal{G} \Sigma_0}{\mathcal{C}_s^2}-\frac{\pi^2 \mathcal{G}^2 \xi R_d \Sigma_0^2}{2 \mathcal{C}_s^4}\,.
\end{equation}
If the RHS of the Eq.~(\ref{disf}) is positive for $k_{\text{min}}$, it will be positive for  all other wavenumbers. Now we substitute $k_{\text{min}}$ from Eq.~(\ref{kmin})  into the RHS of Eq.~(\ref{disf}), and expand the result for small values of $\xi$. Furthermore, considering the definition of the modified Toomre's parameter $\mathcal{Q}=\mathcal{C}_s \kappa/\pi G\Sigma_0$, one can write the stability condition $\omega^2>0$ as follows
\begin{equation}
1-\frac{1}{\mathcal{Q}^2}+\frac{\xi R_d\kappa^2}{2\pi \mathcal{G}\Sigma_0\mathcal{Q}^4}>0.
\end{equation}
This inequality can be rewritten as a condition on the magnitude of $\mathcal{Q}$ 
\begin{equation}
\mathcal{Q}(=\mathcal{Q}_N+\mathcal{A}\mathcal{Q}_c)>1-\frac{\kappa^2 \xi R_d}{4\pi \mathcal{G}\Sigma_0},\label{TC0}
\end{equation}
where, $\mathcal{Q}_N$ ($\mathcal{Q}_c$) is the Newtonian (correction) part of the modified Toomre's parameter\footnote{Note that, we have picked an solution which reproduce the Newtonian criterion after removing the EMSG correction terms. Moreover, the solutions are expanded about $\xi=0$ and $\mathcal{A}=0$ whenever needed.}. It is clear that, removing the small corrections appeared as the coefficients of $\xi$ and $\mathcal{A}$, leads to the standard Toomre's criterion.
Now, regarding to the definitions
$\mathcal{C}_s^2=c_s^2+\alpha^* c^4\Sigma_0$, and $\mathcal{G}=G (1+4\alpha^* c^2 \Sigma_0)$
, and using Eq.~(\ref{kappa}), one can express the new Toomre's criterion (\ref{TC0}) in terms of the Newtonian quantities. As mentioned before, holding this inequality guaranties the stability of the fluid disk against all unstable modes.
In fact, the expanded form of this relation is rather complicated to be written here. However it has been written in the appendix~\ref{app}. 
Finally, the Toomre's criterion in the context of EMSG reads
\begin{equation}
\mathcal{Q}_N>1+\mathcal{C}\left(\Gamma,\mathcal{A},\eta,\mu,y,\xi \right)\,.
\label{TC}  
\end{equation}
Furthermore, $\mathcal{C}\left(\Gamma,\mathcal{A},\eta,\mu,y,\xi \right) $ includes all the corrections introduced by EMSG to the local stability criterion of the rotating gaseous disks. This is one of the most important results of this paper.

Using Eq.~(\ref{TC}), one can compare the stability of the fluid disk in the context of Newtonian gravity and  EMSG. 
In fact, whenever $\mathcal{C}<0$ ($\mathcal{C}>0$), the fluid disk will be more stable (unstable) in the context of EMSG. To see this fact more clearly we refer to the Fig.~\ref{DPlots}.
It should be noted that, assuming $1+\mathcal{C}\left(\Gamma,\mathcal{A},\eta,\mu,y,\xi \right) =RHS$, the stability condition  (\ref{TC}) can be written as $\mathcal{Q}_N-RHS>0$. Therefore, in summary, to compare the Newtonian gravity and the EMSG, we will study the signature of $\mathcal{C}$. Moreover, the pure effects of the parameters in the context of EMSG should be tracked using the stability condition $\mathcal{Q}_N-RHS>0$. These situations are plotted in Fig.~\ref{DPlots}.
The left panel at the first row shows the effect of thickness parameter $\xi$ on the correction term $\mathcal{C}$.
In fact, for a given $\mu$ and $\eta$ parameters, increasing $\xi$ makes this term bigger (but with a minus sign), and so the RHS of Eq.~(\ref{TC}) will be smaller. Therefore, one can see that, increasing $\xi$ leads to stability in system. This is in harmony with the mentioned role for $\beta$ in previous section. Moreover, the left panel at the second row confirms this role as well. In this panel we have shown $\mathcal{Q}_N-RHS$ at radius $y=2$ for fixed values of stability parameters $\eta$ and $\mu$. 
This panel shows that increasing $\xi$, supports the stability of the system. 

To study the effect of $\mathcal{A}$, or equivalently the free parameter $\alpha$ of EMSG, one can see the top right panel in Fig.~\ref{DPlots}. This panel shows that for a positive (negative) value of this parameter, increasing the magnitude of $\mathcal{A}$ makes the disk more stable (unstable). It is clear that, in this situation, the RHS of Eq.~(\ref{TC}) will be smaller (bigger) and the Toomre's criterion will be supported (opposed). This behavior is completely consistent with that explained in Sec.~\ref{5_1} for an infinite medium.  As we saw in the Jeans analysis, a positive (negative) $\alpha$ stabilizes (destabilizes) an infinite non-rotating fluid medium.
This fact also can be seen in the left panel at the second row of Fig.~\ref{DPlots} where we investigated the stability at a fixed radius $y=2$.

It could be also interesting to investigate the role of each parameters $\eta$ and $\mu$ here. As mentioned before, considering Eq.~(\ref{css}), it is clear that, $\mu$ shows the strength of the pressure in fluid disk. Therefore, in general, it is expected for this parameter to  induce stabilizing effects.
For another parameter $\eta$, the situation is more complicated. Let us begin our discussion with looking at Eq.~(\ref{kappa_N}). In some astrophysical systems, in Newtonian regime, where the sound speed is much smaller than the angular velocity $v_{\varphi}=\sqrt{r d\Phi_0/dr}$, the first term in this equation can be ignored. However, it should be noted that, even in the Newtonian viewpoint there are some astrophysical systems like advection-dominated accretion
flows (\cite{Narayan}), where can have $c_s\simeq v_{\varphi}$.
So, the epicycle frequency in this case has a coefficient of $\eta$, and the Toomre's parameter is proportional to $\sqrt{\mu/\eta}$.
Therefore, since $\mu$ has stabilizing effect, it seems that $\eta$, with a destabilizing effect, can be considered to be a representative for the gravity in system. Therefore, these parameters are useful dimensionless quantities to interpret the results and simplify the stability analysis.

The right panel at the second row of Fig.~\ref{DPlots} devoted to studying the role of $\eta$ and $\mu$.
This panel shows that, increasing $\eta$ makes the disk more unstable. It may be expected, because regarding Eq.~(\ref{Epot}), increasing this parameter supports the gravitational strength.
Moreover, this panel shows that, increasing $\mu$ makes the disk more stable. Again, considering Eq.~(\ref{css}), it seems that $\mu$ is a parameter that measures the strength of the pressure in system. Therefore, in general, it is natural to expect such role here. 
It should be noted that, regarding Eq.~(\ref{kappa_N}), for $y<3/2(\Gamma-1)$, the first term could be negative. Therefore, at these radii, a high value of $\mu$ can decrease the magnitude of $\kappa_{\small N}$. As a result, increasing $\mu$, may destabilize the inner radii. This unexpected behaviour of the pressure can be also seen in \cite{kazemi} and \cite{Nazari}. 
Therefore, although these two parameters have been introduced to characterize the role of gravity and pressure, there may be some exceptions that should be treated carefully.

For the sake of completeness, let us study the growth rate of the axisymmetric unstable modes. 
It is not difficult to rewrite Eq.~(\ref{TC}) as
\begin{equation}
\mathcal{S}^2=-q^2(1+\mathcal{H})+\frac{2}{\mathcal{Q}}q-1\,,
\end{equation}
where $\mathcal{S}=i\omega/\kappa$,  $q=\mathcal{C}_s k/\kappa$, and $\mathcal{H}=\pi\mathcal{G}\Sigma_0 \xi R_d/2\mathcal{C}_s^2$. 
\begin{figure*}
	\centering
	\includegraphics[scale=0.5]{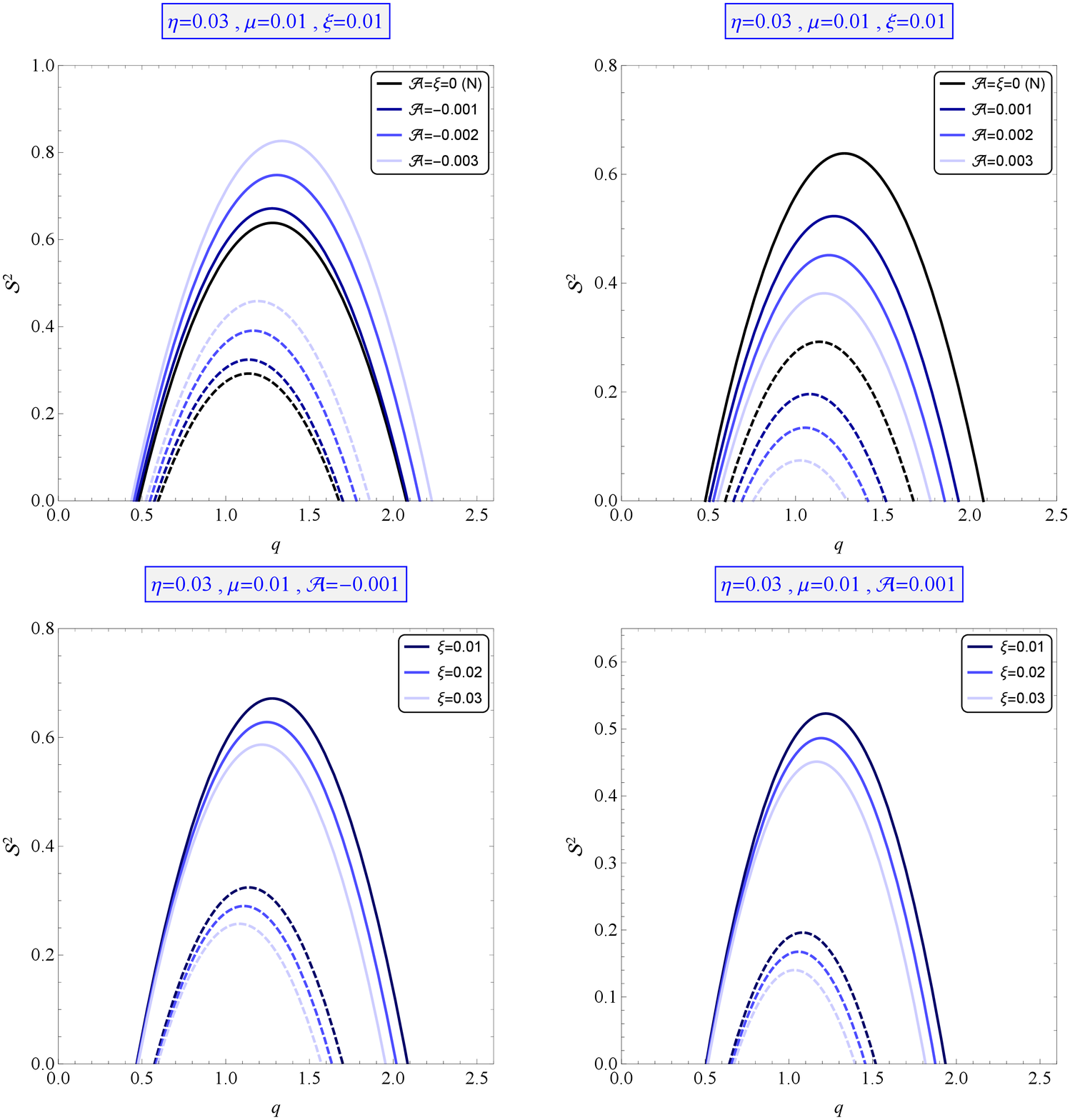}
	\caption{Growth rate of small perturbations in the fluid disk in the context of EMSG.
		It is worth mentioning that, in all panels of growth rate figures, the solid (dashed) curves show $y=1$ ($y=3$) and we have assumed, $\Gamma=2$.
		\label{Rate1}}
\end{figure*}
\begin{figure*}
	\centering
	\includegraphics[scale=0.5]{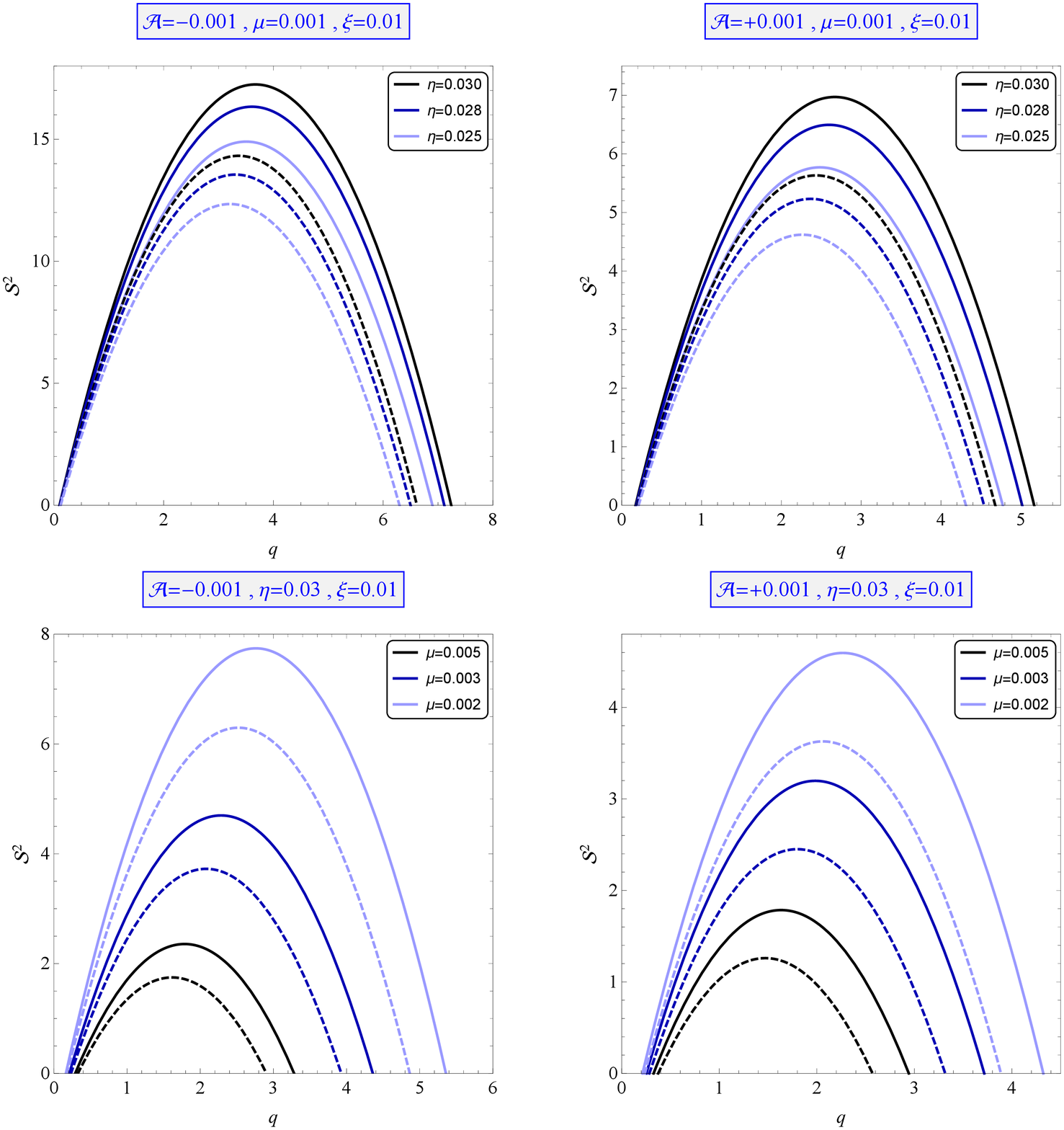}
	\caption{Growth rate of small perturbations in the fluid disk in the context of EMSG.
		\label{Rate2}}
\end{figure*}
\begin{figure*}
	\centering
	\includegraphics[scale=0.5]{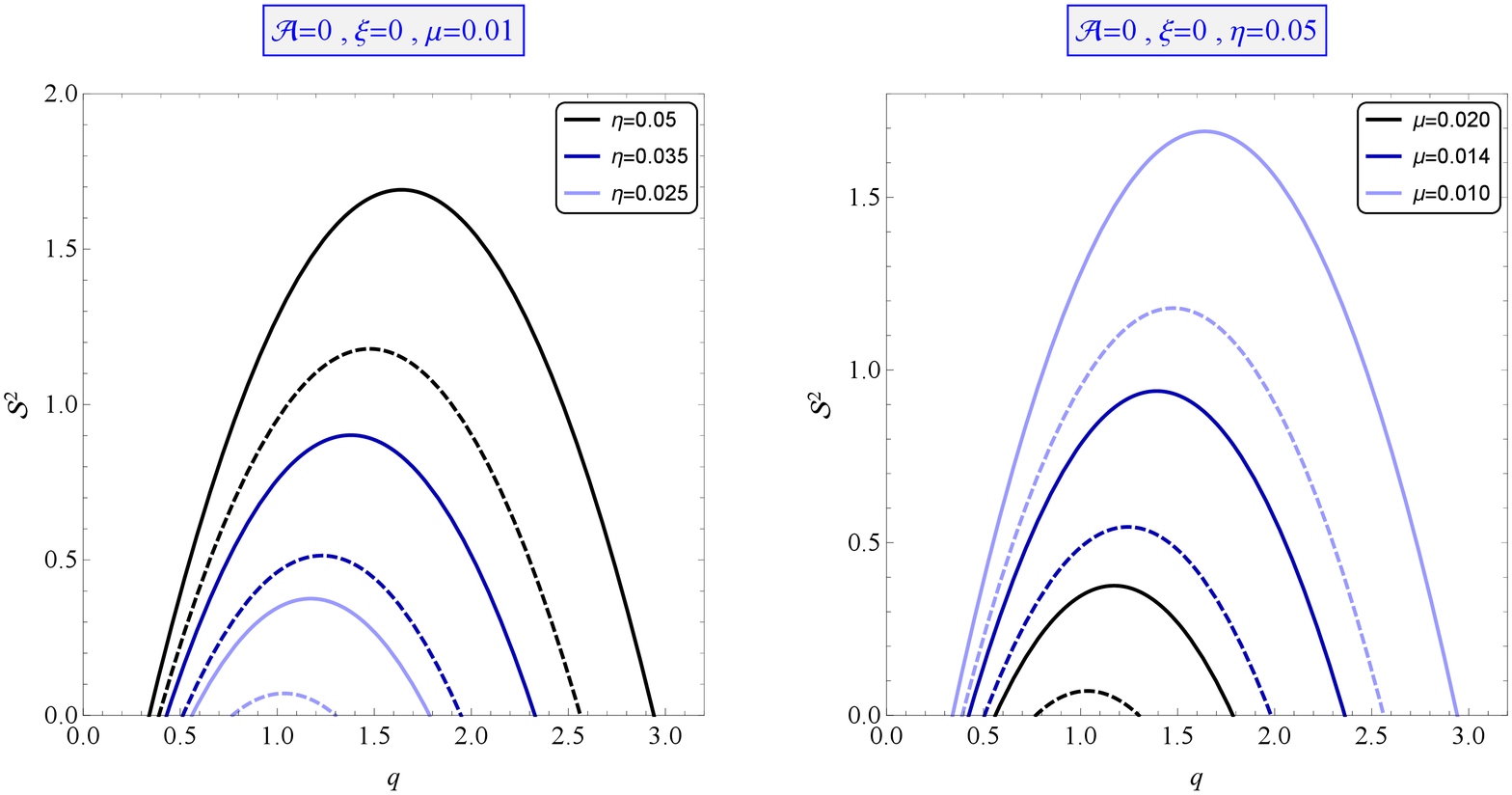}
	\caption{Growth rate of small perturbations in the fluid disk in the context of Newtonian gravity.
		\label{Rate3}}
\end{figure*}
The role of each parameter in the growth rate of the unstable modes for both theories EMSG and Newtonian gravity can be seen in Figs.~\ref{Rate1}-\ref{Rate3}. To have a complete study, all cases are plotted at two different radii. 
The role of the theory's free parameter $\mathcal{A}$ can be found in two top panels of Fig.~\ref{Rate1}. It is clear that, increasing $\mathcal{A}$ decrease the growth rate and consequently stabilizes the disk. This behavior is in harmony with those obtained from Fig.~\ref{DPlots}.
Also, the bottom panels of this figure show that, for both positive and negative values of $\mathcal{A}$, the disk will be stable with the thickness parameter $\xi$.
Moreover, the role of $\eta$ and $\mu$ in the context of EMSG (Newtonian gravity) have been illustrated in the Fig.~\ref{Rate2} (Fig.~\ref{Rate3}).
These figures show that, increasing $\eta$ increases the growth rates. In other words, as one may expected, this parameter makes the disk more unstable in both EMSG and Newtonian gravity.
Also, these figures show that, for both theories, increasing $\mu$, decreases the growth rates and therefore makes the disk more stable. Note that, regarding the mentioned stabilizing role of $\mu$, this behavior was expected. 
Finally, it seems that, the growth rates are higher at the small radii for both theories.
In some senses, it is expected. In fact at central parts we expect high surface density and consequently stronger gravity. Naturally, stronger gravity leads to higher growth rate.

\section{Applying the results to an astrophysical system}\label{applications}

As a realistic system, the new Toomre's criterion can be studied in HMNSs. An HMNS is a resulting object in the merging of a neutron star binary. Because of including the strong gravitational field, and also fast movements, this system seems to be a good candidate to track the footprints of the relativistic effects in local fragmentation.
However, the physical properties of HMNSs are not well known yet because of their complicated evolution.
On the other hand, there are many attempts to study the main properties of HMNSs using numerical simulations in GR and also approximative methods (for example see \cite{Hanauske} and \cite{Hotokezaka}).

In this section, using a toy model  recently introduced in \cite{kazemi} and \cite{RoshanEiBI}, we try to roughly estimate the possibility of local fragmentation in an HMNS. In these studies, the stability parameters $\eta$ and $\mu$ are found for an HMNS using five models of the EOS  studied in \cite{Hanauske}. In fact, using Table 1 of \cite{kazemi} the gravitational local stability has been studied. 
\begin{figure}
	\centering
	\includegraphics[scale=0.7]{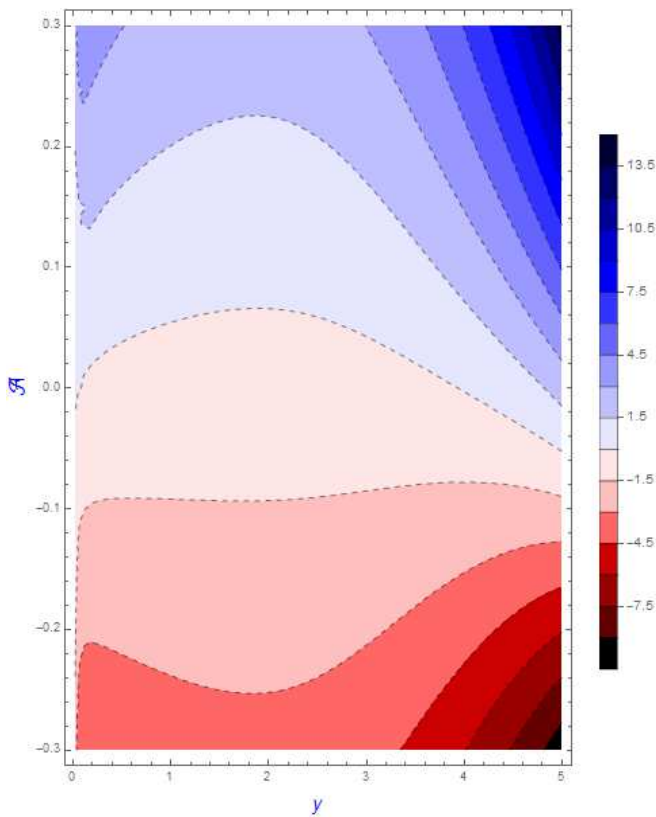}
	\caption{The Toomre's criterion (TC), $\mathcal{Q}-RHS$ versus the dimensionless radius $y$ and the model parameter $\mathcal{A}$ for GNH3-M125 (for more details see the Table 1 of \cite{kazemi}). Here $\Gamma=2$.
		\label{DP}}
\end{figure}
Here, we focus on the model GNH3-M125 only. However, the behavior of the others are more or less similar. First, let us study the case $\mathcal{A}>0$. The Toomre's criterion in the context of EMSG for this case is shown in the top  side of Fig.~\ref{DP}. It is clear that, for a positive value of $\mathcal{A}$, the system will be stable at outer radii. Furthermore, increasing the value of $\mathcal{A}$ in this case leads to a more stable system. In this case, the edge of disk could be the most stable part of the system for a given $\mathcal{A}$. 
On the other hand, as one can see in the bottom  side of Fig~\ref{DP}, for negative values of $\mathcal{A}$, it is possible for (almost all) the system to be unstable.  
It should be noted that, the Toomre's criterion by itself is not enough to conclude about the occurrence of the instability.
In fact, comparing the dynamical time scales of the system can shed some light on the stability problem.
To ensure about the possibility of the occurrence of the instability in a fluid system, one can compare the time scale for the perturbation growth, i.e., $t\propto 1/|\omega|$, with the dynamical time scale of the system (for HMNS it is typically around a mili second).
It is not difficult to show that, for the unstable area in the bottom  side of Fig~\ref{DP}, the perturbation growth timescale for the most unstable wave-number (see Eq.~\ref{kmin}) and the values $R_d=4.105$ (for GNH3-M125), $\xi=0.1$, is $\sim10^{-4} - 10^{-5}$ s.
Therefore, the perturbation growth timescale is smaller than the dynamical timescale of an HMNS ($\sim 10^{-3}$ s). It means that, it may be possible to  occur the local instability in an HMNS system in the context of EMSG.

Although this is a straightforward and simple outcome of our stability analysis, it can put a serious constraint on the viability of EMSG. To the best of our knowledge, no observational evidence has been reported on the existence of local fragmentation in HMNS. On the other hand EMSG with negative $\alpha$ predicts gravitational instability in this system. Naturally, this means that EMSG with positive $\alpha$ is more acceptable from physical point of view.

\section{Discussion and conclusion}\label{conclusions}

In this paper we studied the local gravitational stability of an infinite fluid (Jeans analysis) and also a differentially rotating fluid disk (Toomre's criterion) in the context of EMSG. Firstly, by introducing the field equations of the EMSG and finding the weak field limit of this theory, we derived the modified version of the Poisson's equation. Although, two different cases $f_{RR}\ne0$ and $f_{RR}=0$ can be studied, we only focused on the latter case for practical aims. In fact, the case $f_{RR}\neq0$ is totally reminiscent of the weak field limit of the $f(R)$ gravity. As we already mentioned, in this case because of an inherent non-linearity and consequently screening effects, one cannot simply linearize the field equations, for more details see \cite{hu2007models}. We left the Jeans analysis of this case as a subject for another separate study.

By deriving the hydrodynamics equations and assuming a polytropic EOS we studied the local gravitational stability. An infinite homogeneous self gravitating fluid is the first system which is studied. In this case, by linearizing the hydrodynamics equations as well as the Poisson's equation we found the dispersion relation, and by setting $\omega^2=0$ the Jeans wavenumber could be derived. By achieving the Jeans mass, we showed that the EMSG could have a stabilizing (destabilizing) effect for a positive (negative) value of the model parameter $\alpha$. Moreover, increasing $\alpha$ makes the system more stable for both cases $\alpha>0$ and $\alpha<0$. 

Afterwards, we considered a fluid disk.
Also, to skip some complexities and keep the analysis self-consistent, we have assumed a finite thickness for the disk.
Then, by achieving the potential of a WKB density wave, and also using a perturbative method, we derived the dispersion relation. Finally, defining the modified versions of the sound speed, the gravitational constant, and the epicyclic frequency, the so-called Toomre's criterion is achieved in the context of EMSG. 
Then, considering an exponential surface density profile, and also dimensionless parameters $\eta$ (related to strength of gravity), $\mu$ (related to the pressure in system), and $\mathcal{A}$ (dimensionless model parameter), the modified version of Toomre's criterion can be rewritten in terms of the standard case.
It is interesting that, the general form of this criterion could be written as $\mathcal{Q}_N>1+\mathcal{C}\left(\Gamma,\mathcal{A},\eta,\mu,y,\xi \right)$, where an additional correction term is included here. Again, the EMSG may stabilize or destabilize the disk depending on the sign of the model parameter. However, in both cases, increasing $\mathcal
A$ will support the stability of the system. 
To conduct a more detailed stability analysis, we studied the rate of growing  unstable modes in the disk. We showed that, for both cases $\mathcal{A}>0$ and $\mathcal{A}<0$, increasing $\mathcal{A}$ will makes the disk more stable. Moreover, the growth rate decreases with radius in both EMSG and Newtonian gravity. 

In the last part, using a toy model which has been introduced in \cite{RoshanEiBI} and \cite{kazemi}, we applied our results to an HMNS. We showed that, for a negative value of $\mathcal{A}$ the local fragmentation could be possible in an HMNS. However, a positive $\mathcal{A}$,  in agreement with the observations and numerical simulations, could exclude (some parts of) the system to be locally fragmented. 

\section*{acknowledgment}
This work is supported by Ferdowsi University of Mashhad under Grant NO.  47803, (17/07/1397). This article is based upon work from COST Action CA1511 Cosmology and Astrophysics Network for Theoretical Advances and Training Actions (CANTATA), supported by COST (European Cooperation in Science and Technology).M.D.L. acknowledge INFN Sez. di Napoli (Iniziative Specifiche QGSKY and TEONGRAV). IDM is supported by the grant "The Milky Way and Dwarf Weights with Space Scales" funded by University of Torino and Compagnia di S. Paolo (UniTO-CSP). IDM also
acknowledges partial support from the INFN grant InDark and the Italian Ministry of Education, University and Research (MIUR) under the Departments of Excellence grant L.232/2016

\appendix
\section{Expanded form of the Toomre's criterion in EMSG}\label{app}

Here one can see the expanded form of Eq.~(\ref{TC}). This relation reads
\begin{eqnarray}
\nonumber\mathcal{Q}_N && >1+\mathcal{C}(\Gamma,\mathcal{A},\eta,\mu,y,\xi)\\
&& >1+\xi\mathcal{C}_1+\mathcal{A}\left( -\frac{\mathcal{C}_2}{\mathcal{C}_3}+\frac{\xi}{8}\left( \mathcal{C}_4-\frac{\mathcal{C}_5+\frac{\mathcal{C}_6}{\mathcal{C}_7}}{\mathcal{C}_8}\right) \right) ,
\end{eqnarray}
where
\begin{eqnarray}
\nonumber\mathcal{C}_1  = && \frac{1}{8} \bigg(-\frac{\Gamma  \mu  e^{-2 (\Gamma -2)
		y} (2 (\Gamma -1) y-3)}{\pi  \eta  y}-4 e^{2 y} ((2
I_0(y)\\
&&+y I_1(y)) K_0(y)-(y I_0(y)+I_1(y))
K_1(y))\bigg),
\end{eqnarray}
\begin{eqnarray}
\nonumber\mathcal{C}_2 && =e^{\Gamma  y-(3 \Gamma +1) y} \bigg[2 \pi  \eta 
y \bigg(\left(e^{4 \Gamma  y}-8 \Gamma  \mu  e^{2
	(\Gamma +1) y}\right) \\\nonumber
&&\times ((2 I_0(y)+y I_1(y)) K_0(y)-(y
I_0(y)+I_1(y))\\\nonumber
&&\times  K_1(y))+4 \Gamma  \mu  e^{2 (\Gamma
	+2) y} (2 (I_0(2 y)+ y I_1(2 y))\\\nonumber
&&\times  K_0(2 y)-(2 y I_0(2y)+I_1(2 y)) K_1(2 y))\bigg)+\Gamma  \mu  \\\nonumber
&&\times\Big(e^{2
	(\Gamma +1) y} (\Gamma  y-3)-4 \Gamma  \mu  e^{4 y}(2 (\Gamma -1) y-3)\Big)\bigg], \\
\end{eqnarray}
\begin{eqnarray}
\nonumber\mathcal{C}_3= && \pi\eta  \bigg(8\Gamma\mu y \pi
\eta  y e^{2 \Gamma  y} I_1(y) (y K_0(y)-K_1(y))\\\nonumber
&&-4	\pi  \eta  y e^{2 \Gamma  y} I_0(y) (y K_1(y)-2
K_0(y))+\Gamma  \mu  e^{2 y} \\
&& \times(2 (\Gamma -1) y-3)\bigg)^{1/2},
\end{eqnarray}
\begin{eqnarray}
\nonumber\mathcal{C}_4= && 
\frac{4 \Gamma  \mu  e^{-2 (\Gamma -1) y} (2 (\Gamma -1)
	y-3)}{\pi  \eta  y}+\frac{3-2 y}{\pi  \eta  y}\\\nonumber
&&+16 ((2
I_0(y) +y I_1(y)) K_0(y)-(y I_0(y)+I_1(y)) \\\nonumber
&&\times K_1(y))-16
e^{2 y}  (I_1(2 y) (2 y K_0(2 y)-K_1(2 y))\\
&&+2 I_0(2 y)
(K_0(2 y)-y K_1(2 y)))
\end{eqnarray}
\begin{eqnarray}
\nonumber\mathcal{C}_5= && 
-2 \sqrt{2} e^y \sqrt{y} \bigg(I_1(y) (y K_0(y)-K_1(y))
\Big(e^{2 (\Gamma -1) y}\\\nonumber
&&-8 \Gamma  \mu
\Big)+I_0(y) (2 K_0(y)-y K_1(y)) \Big(e^{2 (\Gamma
	-1) y}\\\nonumber
&&-8 \Gamma  \mu \Big)+8 \Gamma  \mu  e^{2 y}(I_1(2 y) (2 y K_0(2 y)-K_1(2 y))\\
&&+2 I_0(2 y) (K_0(2
y)-y K_1(2 y)))\bigg)
\end{eqnarray}
\begin{eqnarray}
\nonumber\mathcal{C}_6=&&
2 \sqrt{2} \sqrt{y} e^{2 \Gamma  y-2 (\Gamma +1) y+y}
(I_1(y) (y K_0(y)-K_1(y))\\\nonumber
&&+I_0(y) (2 K_0(y)-y K_1(y)))
\bigg[2 \pi  \eta  y \bigg(\Big(e^{4 \Gamma  y} -8
\Gamma  \mu\\\nonumber
&& \times  e^{2 (\Gamma +1) y}\Big)((2 I_0(y)+y
I_1(y)) K_0(y)-(y I_0(y)\\\nonumber
&& +I_1(y)) K_1(y))+4 \Gamma 
\mu  e^{2 (\Gamma +2) y} (2 (I_0(2 y)\\\nonumber
&&  +y I_1(2 y))
K_0(2 y)-(2 y I_0(2 y)+I_1(2 y)) K_1(2
y))\bigg) \\\nonumber
&&+\Gamma  \mu \bigg(e^{2 (\Gamma +1) y}
(\Gamma  y-3)-4 \Gamma  \mu  e^{4 y} (2 (\Gamma -1)
y\\
&&-3)\bigg)\bigg]
\end{eqnarray}
\begin{eqnarray}
\nonumber\mathcal{C}_7= && 
4 \pi  \eta  y e^{2 \Gamma  y} I_1(y) (y
K_0(y)-K_1(y))-4 \pi  \eta  y e^{2 \Gamma  y} I_0(y)\\\nonumber
&&
\times(y K_1(y)-2 K_0(y))+\Gamma  \mu  e^{2 y} (2 (\Gamma
-1) y-3)\\
\end{eqnarray}
\begin{eqnarray}
\nonumber\mathcal{C}_8 && = 
\Big((\mu\Gamma )(4 \pi  \eta  y e^{2
	\Gamma  y} I_1(y) (y K_0(y)-K_1(y)) \\\nonumber
&& -4 \pi  \eta  y
e^{2 \Gamma  y} I_0(y)(y K_1(y)-2 K_0(y))+\Gamma 
\mu  e^{2 y}\\
&&\times (2 (\Gamma -1) y-3))\Big)^{1/2}
\end{eqnarray}
Again, it is clear that, regarding this equation, ignoring the coefficients of $\xi$ and $\mathcal{A}$, the Standard Toomre's criterion $\mathcal{Q}_N>1$ will be reproduced.


\bibliographystyle{spphys}       
\bibliography{EMSGJeans}   

\providecommand{\noopsort}[1]{}\providecommand{\singleletter}[1]{#1}%
\begin{thebibliography}{10}
\providecommand{\url}[1]{{#1}}
\providecommand{\urlprefix}{URL }
\expandafter\ifx\csname urlstyle\endcsname\relax
  \providecommand{\doi}[1]{DOI \discretionary{}{}{}#1}\else
  \providecommand{\doi}{DOI \discretionary{}{}{}\begingroup
  \urlstyle{rm}\Url}\fi

\bibitem{Blake2011}
C.~Blake, E.A. Kazin, F.~Beutler, T.M. Davis, D.~Parkinson, S.~Brough,
  M.~Colless, C.~Contreras, W.~Couch, S.~Croom, et~al., Mon. Not. R. Astron.
  Soc. \textbf{418}(3), 1707 (2011)

\bibitem{Suzuki2012}
N.~Suzuki, D.~Rubin, C.~Lidman, G.~Aldering, R.~Amanullah, K.~Barbary,
  L.~Barrientos, J.~Botyanszki, M.~Brodwin, N.~Connolly, et~al., Astrophys. J.
  \textbf{746}(1), 85 (2012)

\bibitem{Hinshaw2013}
G.~Hinshaw, D.~Larson, E.~Komatsu, D.~Spergel, C.~Bennett, J.~Dunkley,
  M.~Nolta, M.~Halpern, R.~Hill, N.~Odegard, et~al., Astrophys. J. Suppl. Ser.
  \textbf{208}(2), 19 (2013)

\bibitem{Planck18}
N.~Aghanim, Y.~Akrami, M.~Ashdown, J.~Aumont, C.~Baccigalupi, M.~Ballardini,
  A.~Banday, R.~Barreiro, N.~Bartolo, S.~Basak, et~al., arXiv preprint
  arXiv:1807.06209  (2018)

\bibitem{bertone2005}
G.~Bertone, D.~Hooper, J.~Silk, Physics reports \textbf{405}(5-6), 279 (2005)

\bibitem{caldwell2009}
R.R. Caldwell, M.~Kamionkowski, Annual Review of Nuclear and Particle Science
  \textbf{59}, 397 (2009)

\bibitem{Feng2010}
J.L. Feng, Annual Review of Astronomy and Astrophysics \textbf{48}, 495 (2010)

\bibitem{atrio2016}
B.~Wang, E.~Abdalla, F.~Atrio-Barandela, D.~Pavon, Rep. Prog. Phys.
  \textbf{79}(9), 096901 (2016)

\bibitem{PR03}
P.~Peebles, B.~Ratra, Rev. Mod. Phys. \textbf{75}(2), 559 (2003)

\bibitem{Pad03}
T.~Padmanabhan, Physics Reports \textbf{380}(5-6), 235 (2003)

\bibitem{D+05}
M.~Demianski, E.~Piedipalumbo, C.~Rubano, C.~Tortora, Astron. Astrophys.
  \textbf{431}(1), 27 (2005)

\bibitem{CTTC06}
V.F. Cardone, C.~Tortora, A.~Troisi, S.~Capozziello, Phys. Rev. D
  \textbf{73}(4), 043508 (2006)

\bibitem{Caldwell02}
R.R. Caldwell, Phys. Lett. B \textbf{545}(1-2), 23 (2002)

\bibitem{PR88}
P.~Peebles, B.~Ratra, Astrophys. J. \textbf{325}, L17 (1988)

\bibitem{RP88}
B.~Ratra, P.J. Peebles, Phys. Rev. D \textbf{37}(12), 3406 (1988)

\bibitem{SS00}
V.~Sahni, A.~Starobinsky, International Journal of Modern Physics D
  \textbf{9}(04), 373 (2000)

\bibitem{Schive2014}
H.Y. Schive, T.~Chiueh, T.~Broadhurst, Nature Physics \textbf{10}(7), 496
  (2014)

\bibitem{Capolupo2016}
A.~Capolupo, Advances in High Energy Physics \textbf{2016} (2016)

\bibitem{Capolupo2017}
A.~Capolupo, Advances in High Energy Physics \textbf{2018} (2018)

\bibitem{Kleidis2011}
K.~Kleidis, N.K. Spyrou, Astron. Astrophys. \textbf{529}, A26 (2011)

\bibitem{Kleidis2015}
K.~Kleidis, N.~Spyrou, Astron. Astrophys. \textbf{576}, A23 (2015)

\bibitem{Kleidis2017}
K.~Kleidis, N.K. Spyrou, Astron. Astrophys. \textbf{606}, A116 (2017)

\bibitem{demartino2017b}
I.~De~Martino, T.~Broadhurst, S.H.H. Tye, T.~Chiueh, H.Y. Schive, R.~Lazkoz,
  Phys. Rev. Lett. \textbf{119}(22), 221103 (2017)

\bibitem{demartino2018}
I.~De~Martino, T.~Broadhurst, S.H.H. Tye, T.~Chiueh, H.Y. Schive, R.~Lazkoz,
  Galaxies \textbf{6}(1), 10 (2018)

\bibitem{capolupo2019}
A.~Capolupo, I.~De~Martino, G.~Lambiase, A.~Stabile, Phys. Lett. B
  \textbf{790}, 427 (2019)

\bibitem{Misner1970}
C.~Misner.
\newblock Relativity ed m carmeli, s fickler and l witten (1970)

\bibitem{Itzykson1980}
C.~Itzykson, J.B. Zuber, \emph{Quantum field theory} (Courier Corporation,
  2012)

\bibitem{Isham1981}
C.J. Isham, R.~Penrose, D.W. Sciama, \emph{Quantum Gravity 2: A Second Oxford
  Symposium} (Oxford University Press, USA, 1981)

\bibitem{Faraoni2009}
S.~Capozziello, M.~De~Laurentis, V.~Faraoni, The Open Astronomy Journal
  \textbf{3}, 49 (2010)

\bibitem{darkmetric}
S.~Capozziello, M.~De~Laurentis, M.~Francaviglia, S.~Mercadante, Found. Phys.
  \textbf{39}(10), 1161 (2009)

\bibitem{Nojiri2011}
S.~Nojiri, S.D. Odintsov, Physics Reports \textbf{505}(2-4), 59 (2011)

\bibitem{PhysRept}
S.~Capozziello, M.~De~Laurentis, Physics Reports \textbf{509}(4-5), 167 (2011)

\bibitem{Annalen2012}
S.~Capozziello, M.~De~Laurentis, Ann. Phys. \textbf{524}, 545 (2012)

\bibitem{idm2015}
I.~de~Martino, M.~De~Laurentis, S.~Capozziello, Universe \textbf{1}(2), 123
  (2015).
\newblock \doi{10.3390/universe1020123}

\bibitem{Nojiri2017}
S.~Nojiri, S.~Odintsov, V.~Oikonomou, Physics Reports \textbf{692}, 1 (2017)

\bibitem{Nojiri:2006ri}
S.~Nojiri, S.D. Odintsov, International Journal of Geometric Methods in Modern
  Physics \textbf{4}(01), 115 (2007)

\bibitem{Cai2016}
Y.F. Cai, S.~Capozziello, M.~De~Laurentis, E.N. Saridakis, Rep. Prog. Phys.
  \textbf{79}(10), 106901 (2016)

\bibitem{Will93}
C.M. Will, \emph{Theory and experiment in gravitational physics} (Cambridge
  university press, 2018)

\bibitem{Stairs2003}
I.H. Stairs, Living Reviews in Relativity \textbf{6}(1), 5 (2003)

\bibitem{Everitt2011}
C.F. Everitt, D.~DeBra, B.~Parkinson, J.~Turneaure, J.~Conklin, M.~Heifetz,
  G.~Keiser, A.~Silbergleit, T.~Holmes, J.~Kolodziejczak, et~al., Phys. Rev.
  Lett. \textbf{106}(22), 221101 (2011)

\bibitem{us}
M.~Roshan, F.~Shojai, Phys. Rev. D \textbf{94}(4), 044002 (2016)

\bibitem{katirci2014f}
N.~Kat{\i}rc{\i}, M.~Kavuk, Eur. Phys. J. Plus \textbf{129}(8), 163 (2014)

\bibitem{nari}
N.~Nari, M.~Roshan, Phys. Rev. D \textbf{98}(2), 024031 (2018)

\bibitem{Demorest:2010bx}
P.~Demorest, J.~Hessels, T.~Pennucci, S.~Ransom, M.~Roberts, Nature
  \textbf{467}(arXiv: 1010.5788), 1081 (2010)

\bibitem{Antoniadis:2013pzd}
J.~Antoniadis, P.C. Freire, N.~Wex, T.M. Tauris, R.S. Lynch, M.H. van Kerkwijk,
  M.~Kramer, C.~Bassa, V.S. Dhillon, T.~Driebe, et~al., Science
  \textbf{340}(6131), 1233232 (2013)

\bibitem{Binney}
J.~Binney, S.~Tremaine, \emph{Galactic dynamics}, vol.~20 (Princeton university
  press, 2008)

\bibitem{Capozziello2012}
S.~Capozziello, M.~De~Laurentis, I.~De~Martino, M.~Formisano, S.~Odintsov,
  Phys. Rev. D \textbf{85}(4), 044022 (2012)

\bibitem{idm2017a}
I.~De~Martino, A.~Capolupo, The European Physical Journal C \textbf{77}(10),
  715 (2017)

\bibitem{Arbuzova2014}
E.V. Arbuzova, A.D. Dolgov, L.~Reverberi, Phys. Lett. B \textbf{739}, 279
  (2014)

\bibitem{Roshan2014}
M.~Roshan, S.~Abbassi, Phys. Rev. D \textbf{90}, 044010 (2014)

\bibitem{Toomre1964}
A.~Toomre, Astrophys. J. \textbf{139}, 1217 (1964)

\bibitem{Roshan2015a}
M.~Roshan, S.~Abbassi, Astrophys. J. \textbf{802}(1), 9 (2015)

\bibitem{Roshan2015b}
M.~Roshan, S.~Abbassi, Astrophysics and Space Science \textbf{358}(1), 11
  (2015)

\bibitem{RoshanEiBI}
M.~Roshan, A.~Kazemi, I.~De~Martino, Mon. Not. R. Astron. Soc.  (2018)

\bibitem{barrow}
C.V. Board, J.D. Barrow, Phys. Rev. D \textbf{96}(12), 123517 (2017)

\bibitem{akarsu2018cosmic}
{\"O}.~Akarsu, N.~Kat{\i}rc{\i}, S.~Kumar, Phys. Rev. D \textbf{97}(2), 024011
  (2018)

\bibitem{Poisson2013}
E.~Poisson, C.M. Will, \emph{Gravity: Newtonian, Post-Newtonian, Relativistic}
  (Cambridge university press, 2013)

\bibitem{ehlers}
J.~Ehlers, I.~Ozsvath, E.L. Sch{\"u}cking, Y.~Shang, Phys. Rev. D
  \textbf{72}(12), 124003 (2005)

\bibitem{Koivisto}
T.~Koivisto, Classical and Quantum Gravity \textbf{23}(12), 4289 (2006)

\bibitem{kazemi}
A.~Kazemi, M.~Roshan, E.~Nazari, Astrophys. J. \textbf{865}(1), 71 (2018)

\bibitem{hu2007models}
W.~Hu, I.~Sawicki, Phys. Rev. D \textbf{76}(6), 064004 (2007)

\bibitem{1984ApJ...276..114J}
C.J. Jog, P.~Solomon, Astrophys. J. \textbf{276}, 114 (1984)

\bibitem{1970ApJ...161...87V}
P.O. Vandervoort, Astrophys. J. \textbf{161}, 87 (1970)

\bibitem{1992MNRAS.256..307R}
A.B. Romeo, Mon. Not. R. Astron. Soc. \textbf{256}(2), 307 (1992)

\bibitem{2007tisp.book.....G}
I.S. {Gradshteyn}, I.M. {Ryzhik}, A.~{Jeffrey}, D.~{Zwillinger}, \emph{{Table
  of Integrals, Series, and Products}} (2007)

\bibitem{Narayan}
R.~Narayan, I.s. Yi, Astrophys. J. \textbf{428}, L13 (1994)

\bibitem{Nazari}
E.~Nazari, A.~Kazemi, M.~Roshan, S.~Abbassi, Astrophys. J. \textbf{839}(2), 75
  (2017)

\bibitem{Hanauske}
M.~Hanauske, K.~Takami, L.~Bovard, L.~Rezzolla, J.A. Font, F.~Galeazzi,
  H.~St{\"o}cker, Phys. Rev. D \textbf{96}(4), 043004 (2017)

\bibitem{Hotokezaka}
K.~Hotokezaka, K.~Kiuchi, K.~Kyutoku, T.~Muranushi, Y.i. Sekiguchi, M.~Shibata,
  K.~Taniguchi, Phys. Rev. D \textbf{88}(4), 044026 (2013)

\end{thebibliography}

%
%

\end{document}